%% file: grid-peeling.tex
\newif\ifArxiv
\Arxivtrue

\ifArxiv
\documentclass[11pt]{article}
\usepackage[a4paper,margin=25mm]{geometry}
\else
\documentclass[a4paper,
cleveref, autoref
]{socg-lipics-v2021}
\fi

\usepackage{graphicx} 
\usepackage{amsmath,amsthm,amsfonts,xcolor}
\allowdisplaybreaks[1]
\usepackage{hyperref}
\usepackage{enumerate}
\newif \iflong
\longfalse 
\ifArxiv\longtrue\fi
\newif \ifanonymous

\ifArxiv
 \newtheorem{lemma}{Lemma}

 \newtheorem{theorem}{Theorem}
 \newtheorem{proposition}{Proposition}
 \newtheorem{observation}{Observation}

 \theoremstyle{definition}
 \newtheorem{definition}{Definition}
\fi

\newcommand{\eps}{\varepsilon}

\DeclareMathOperator{\lcm}{lcm}

\title{Grid Peeling of Parabolas}

\ifArxiv
\author{G\"unter Rote, Moritz R\"uber, Morteza Saghafian}

\else
\titlerunning{Grid peeling of parabolas}

\EventEditors{Wolfgang Mulzer and Jeff M. Phillips}
\EventNoEds{2}
\EventLongTitle{40th International Symposium on Computational Geometry (SoCG 2024)}
\EventShortTitle{SoCG 2024}
\EventAcronym{SoCG}
\EventYear{2024}
\EventDate{June 11--14, 2024}
\EventLocation{Athens, Greece}
\EventLogo{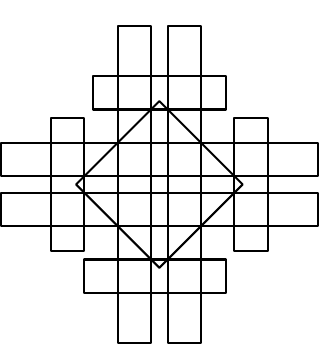}
\SeriesVolume{293}
\ArticleNo{XX}     

\ifanonymous
\author{Blind Author(s)}{\ }{}{}{}
\authorrunning{Grid peeling of parabolas}
\Copyright{Double-Blind Author(s)}
\EventShortTitle{for SoCG 2024}
\EventAcronym{SoCG}
\EventYear{2024}

\ArticleNo{115}
\else

\author{G\"unter Rote}{Freie Universit\"at Berlin}{rote@inf.fu-berlin.de}{https://orcid.org/0000-0002-0351-5945}{}

\author{Moritz R\"uber}{Freie Universit\"at Berlin}{m.rueber@web.de}{}{}

\author{Morteza Saghafian}{ISTA (Institute of Science and Technology Austria), Kloster\-neu\-burg, Austria}{Morteza.Saghafian@ist.ac.at}{https://orcid.org/0000-0002-4201-5775}{}

\authorrunning{G. Rote, M. R\"uber, and M. Saghafian}

\Copyright{G\"unter Rote, Moritz R\"uber, and Morteza Saghafian}

\relatedversion{A full version of the paper is available at arXiv:\href{https://arxiv.org/abs/papernumber}{2402.999}.} 

\fi


\fi

\begin{document}

\maketitle

\begin{abstract}
 Grid peeling is the process of repeatedly removing the
 convex hull vertices of the grid-points that lie inside a given
 convex curve.
 It has been conjectured that, for a more and more refined grid,
 grid peeling converges to a continuous process, the
 \emph{affine curve-shortening flow}, which deforms the curve
 based on the curvature.
 \iflong

 \fi
 We prove this conjecture for one class of curves, parabolas with a vertical axis, and we determine the value of the constant factor in the formula that relates the two processes.
\end{abstract}
\iflong
\tableofcontents
\fi
\section{Introduction}

In 2017, Eppstein, Har-Peled, and Nivasch
\cite{ehn-gpacs-20} observed a remarkable connection between a
continuous deformation of a curve in the plane, the 
\emph{affine curve-shortening flow} (ACSF), and a discrete process,
\emph{grid peeling}.

\begin{figure}[htb]
    \centering
    \includegraphics[scale=0.9]{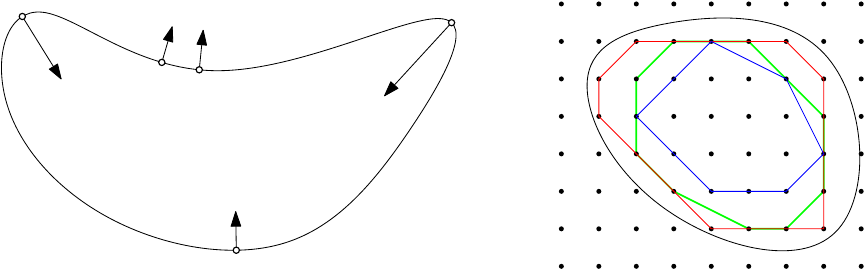}
    \caption{
  Left: The Affine Curve-
 Shortening Flow (ACSF).
 The velocity 
 is indicated by
 arrows, whose length is proportional to $\kappa^{1/3}$.
 Right: A convex curve and the first three steps of grid peeling.
 }
    \label{fig:ACSF}
\end{figure}

\begin{figure}
    \centering
    \includegraphics[scale=0.9]{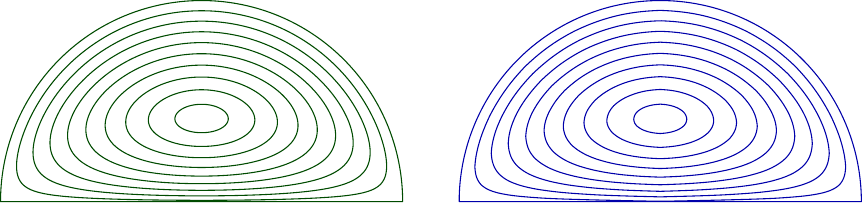}
    \caption{ACSF (left) and grid peeling (right) of a semicircle
    of diameter~1.
The left figure shows 10 snapshots of ACSF with regular time increments
;
the right figure shows every 2714th convex layer for a grid of spacing $1/5000$.
\iflong
The increment 2714 corresponds to
the conjectured formula \eqref{m_conj}
with a value $c_{\mathrm{g}}\approx 1.587$.
%
\fi
    (From \cite[Figure~3]{ehn-gpacs-20}, by permission from the authors.)}
    \label{fig:semicircle}
\end{figure}

In the affine curve-shortening flow, a smooth curve is deformed by moving every point
toward the direction in which the curve bends, 
at a speed of $\kappa^{1/3}$ in the normal direction,
where $\kappa$ is the curvature at that point at the
current point of time, see
Figure~\ref{fig:ACSF}. 
\iflong 
The left part of \fi
Figure~\ref{fig:semicircle} shows
\iflong a few \fi snapshots of this
inward-growing
process, starting from a semicircle.
(The semicircle is not a smooth curve, but the definition
of the flow
can be extended to cover
\iflong piecewise smooth \else such \fi curves.)

By contrast, grid peeling is a process that is discrete both in space and in time.
Given a convex curve, we start by finding the convex hull of all
points
of a uniform square grid inside the curve.
Then we iteratively remove the vertices of the convex hull, and take
the convex hull of the remaining grid points.

Eppstein, Har-Peled, and Nivasch observed
that, as the underlying grid is refined, grid peeling approximates
the ACSF process, as can be seen in
 Figure~\ref{fig:semicircle}.
More specifically\iflong\else\ 
\cite[Conjecture~1]{ehn-gpacs-20}\fi, they conjectured
that, for a more and more refined grid of spacing $1/n$, the $m$-th
convex layer of a convex curve converges to
the ACSF after time $T$,
if $m$ is chosen as
\begin{equation}\label{m_conj}
m = \lfloor c_{\mathrm{g}} T n^{4/3} \rfloor
\end{equation}
for an appropriate constant $c_{\mathrm{g}}$, which was experimentally determined
to be approximately~
$1.6$\iflong
\ \cite[Conjecture~1]{ehn-gpacs-20}\fi.

We prove this conjecture for the special case when the
curve is a parabola
with vertical 
axis,
and we find the precise value of the constant $c_{\mathrm{g}}$:
\begin{equation}
    \label{Cg}
c_{\mathrm{g}} = \sqrt[3]{\frac {\pi^2}{2\zeta(3)}}\approx 1.60120980542577,
\end{equation}
where
\iflong
  \begin{displaymath}
\else $\fi
    \zeta(s) = 1+\frac 1{2^s}+\frac 1{3^s}+\frac 1{4^s}+\cdots
\iflong    
  \end{displaymath}
\else $ \fi
  is the Riemann zeta function\iflong, with values
  $\zeta(2)=\pi^2/6\approx 1.644934$ and $\zeta(3)\approx 1.2020569$\fi.

\begin{theorem}\label{main-parabola}
   For the parabola
   $\Pi\colon y=ax^2/2 + bx + c$,
the ACSF is
a vertical translation with velocity
 $a^{1/3}$.
 Thus, at time $T>0$, it becomes the
 parabola $\Pi^T\colon y=ax^2/2 + bx + c+ Ta^{1/3}$.
 
If we apply grid peeling to $\Pi$
with a grid of spacing $1/n$
for
$m = \lfloor c_{\mathrm{g}} T n^{4/3} \rfloor$
steps, then, as $n\to\infty$,
the vertical distance between the resulting
grid polygon and $\Pi^T$ is
bounded by
\begin{equation}
    \nonumber
    O\biggl(\frac{(Ta^{2/3}+a^{-2/3})\log \frac na}{n^{1/3}} \biggr).
\end{equation}
\end{theorem}
 For fixed $T$ and $a$, this error bound goes to $0$ as $n\to \infty$.

We 
can extend this theorem to any parabola whose axis has
a rational slope $a/b$: A~uni\-modular transformation with a suitable matrix
$\left(\begin{smallmatrix}
a&-b\\
u&v\\
\end{smallmatrix}
\right)$
of determinant 1
will leave the grid unchanged and make the axis vertical, and then Theorem~\ref{main-parabola} can be applied.

\subsection{History and background}

The ACSF process was
  first studied
in the 1990s in the area of computer vision and image processing,
by Alvarez, 
 Guichard, 
 Lions, 
and Morel~\cite{alv93} and by 
Sapiro and Tannenbaum~\cite{sap93}.
One way to understand the ACSF is to regard it
as a limit of \emph{affine erosions}, as shown by
F.~Cao~\cite[
Theorem~6.22]{Cao2003}.
An affine erosion with parameter $\eps$
removes the union of all pieces of area $\eps$ that can be cut off
by a straight line.
(In convex geometry, this is
also called the \emph{wet part}; it plays
a role in 
estimating the area and the number of vertices of 
the convex hull of
a random sample of points 
\cite{BarLar,bfghr-rpwpa-20}.)
Repeating this process makes the shape
rounder and rounder,
like a pebble rolling in water.
Letting $\eps$ go to zero leads to the ACSF as
the continuous limit.
\iflong\else\looseness-1\fi

The other process that we study is formed by
the \emph{convex layers} or \emph{onion layers} of a point set.
They have
their origin in computational geometry and statistics: The innermost
convex layer provides a robust estimate of the ``center'' of a
distribution.
The special case where the point set is a grid was
first investigated 
by 
Har-Peled and Lidický
\cite{Har-Peled-Lidicky}, who showed that
the 
$n\times n$ square grid
has $\Theta(n ^{4/3})$ convex layers. 
For a box in three and higher dimensions, the asymptotic number of layers is
not known,
see \iflong for example \fi
\cite{Dillon-Varada-2023} and the references given there.
See \cite{homotopic-avva-21,cs-lschp-20
,Cao2003}
for more background and references to the literature, both on the ACSF and on grid peeling.

\subsubsection{Peeling with random sets.}
More recently, Calder and Smart
\cite{cs-lschp-20}
investigated the related process where the
grid is replaced by
a random point set.
More precisely, the refined
grid of spacing $1/n$ is replaced by
a Poisson point set of density $1/n^2$.
In this setting, they could prove
an analogous statement:

There exists a constant $c_{\mathrm{r}}\approx 1.3$
%
%
%
%
%
%
such that
the $m$-th convex layer, for
$m = \lfloor c_{\mathrm{r}} T n^{4/3} \rfloor$,
approximates the ACSF at time~$T$.
Since the underlying process is random,
this statement requires some probabilistic qualification;
see
\cite[Theorem~1.2]{cs-lschp-20}
for the precise statement, which is quite strong and general:
It is valid in arbitrary dimension, and
convergence holds (with high probability)
uniformly for all $T$. 
\iflong
The density can be nonuniform,
which corresponds to an ACSF with a
location-sensitive speed.
There is no precise formula
for the value of
the random-set peeling constant $c_{\mathrm{r}}$,
 not even a conjectured one.
Since $c_{\mathrm{r}}<c_{\mathrm{g}}$, random-set peeling proceeds
faster than grid peeling at 
the same 
density.
\fi

\subsubsection{Homotopic curve shortening.}
Avvakumov and Nivasch \cite{homotopic-avva-21} 
extended 
peeling 
to nonconvex and even self-crossing curves,
introducing the concept of \emph{homotopic curve shortening}.
Both for grid peeling and for random-set peeling,
the observed 
relation with the ACSF process persists also in this setting.
\iflong\else\looseness-1\fi

\subsubsection{Equivariance under affine transformations.}
It is easy to check that
the 
ACSF is equivariant under 
area-preserving affine transformations,
a property that gave rise to the term
``\emph{affine} curve-shortening flow.''
\iflong
(Arbitrary affine transformations, which are not
necessarily area-preserving, can be accommodated by
scaling the time parameter.)
\fi
The relation between ACSF and grid peeling is the more surprising as
grid peeling 
\iflong
does not have this property.
Grid peeling 
\fi
is equivariant only under
a special class of affine transformations, namely those
that also preserve the grid (unimodular transformations),
a property that we will often use.
(Peeling with random sets, on the other hand, is
clearly 
equivariant under
area-preserving affine transformations.)

\subsection{Conics}
\label{sec:hyperbolas}
 As stated in Theorem~\ref{main-parabola},
 the ACSF for a parabola is 
 just 
 a translation at
 constant speed.
 This special behavior is shared, to a certain extent,
by the other types of conics: They are scaled under ACSF
but otherwise maintain their shape
\cite[Lemma 8]{sap-tan-1994}.
More specifically,
\begin{itemize}
\item an ellipse (or a circle) \emph{shrinks} toward the center, and eventually collapses to a point;
\item a parabola is \emph{translated} parallel to the axis;
\item a hyperbola \emph{expands} from its center.
\end{itemize}
Among the conics,
parabolas appear
most attractive for investigation, because they don't
even need to be scaled.
Also for the case of random-set peeling,
the peeling of a parabola lies at the core of the proof of
Calder and Smart~\cite{cs-lschp-20}, forming what they call the \emph{cell problem}.
As regards experiments,
the downside of parabolas, as opposed to ellipses,
is that a parabola is an unbounded curve,
and even the first step of grid peeling
is not obvious to compute.
However, as we shall see,
for parabolas with rational coefficients, we can make use
of a certain periodicity 
along the curve, which reduces grid peeling
to a finite computation.
Once the sequence of peelings
goes into a loop, one has a complete overview
of the whole infinite 
\iflong
grid peeling
process.
\else
process.
\looseness-1
\fi

Grid peeling has been investigated also for \emph{hyperbolas}, in a sense.
If one starts with the upper-right quadrant $\mathbb{R}_+
\times \mathbb{R}_+$, the ACSF develops into
positive branches of hyperbolas $xy=c$. 
Eppstein, Har-Peled, and Nivasch
\cite[Theorem 5]{ehn-gpacs-20} investigated the convex
layers of $\mathbb N\times\mathbb N
$
and proved that
the $m$-th convex layer 
is sandwiched between two hyperbolas:
\begin{equation}
    \label{squeeze}
c_1 m^{3/2} \le xy \le c_2 m^{3/2},
\end{equation}
except that the lower bound does not hold 
\iflong
within
a strip of width 
$O(\sqrt m\log^2 m)$ around 
the axes.

The constants $c_1$ and $c_2$ are not computed explicitly, but some values can be worked out
from the proof.
We add two side remarks regarding the exception near the axes that
the theorem makes. Firstly, \emph{some} exception of this sort has to be made, because the $m$-th layer
goes through the points $(m,0)$ and $(0,m)$, and
no hyperbola $xy=\mathrm{const}$ can be squeezed below these points.
To give the hyperbola some chance in principle to squeeze under, we might
peel the quadrant of \emph{positive} grid points, or equivalently, we allow 
the hyperbola to be centered at $(-1,-1)$ (or some other fixed point), but this
would still not suffice for~\eqref{squeeze}.
Secondly,
the claim
\cite[Theorem 5]{ehn-gpacs-20} is stated with an ``exception
strip'' around the axes whose width is only
$O(\sqrt m)$; however, there is a small gap in the proof
\cite[p.~315, right column]{ehn-gpacs-20}:
In the proof of Lemma 18,
when applying Lemma 7 
for the rectangle spanned by the points $q/2$ and $q$,
an error term of the form $\pm O(N \log N)$ from Lemma 7 is ignored.
When the error term is taken into account, the proof goes through with the larger margin of exception claimed above.
\else
within a distance
$O(\sqrt m\log^2 m)$ from
the axes.\footnote
{\cite[Theorem 5]{ehn-gpacs-20} claims an ``exception strip''  around the axes with a smaller width
of $O(\sqrt m)$; however, in the proof of that theorem,
an error term of the form $\pm O(N \log N)$ from Lemma 7 is ignored.}
\fi

\subsection{Overview}

In Section~\ref{sec:grid-parabola}, we define a family of specific
curves, 
the so-called \emph{grid parabolas}~$P_t$.
Grid peeling reproduces them with a vertical shift after $t$ steps
(or $t+1$, depending on the parity).
This is our main technical result (Theorem~\ref{th:reproduce}),
whose proof is postponed to
Section~\ref{sec:proof-grid}.
Based on Theorem~\ref{th:reproduce}, we prove Theorem~\ref{main-parabola}, our main theorem about grid peeling of parabolas,
in
Section~\ref{sec:main-result}.
\iflong
Sections
\else
Appendix
\fi
\ref{Ht-alternative} and~\ref{proof-alpha}
analyze the quantities that arise in the construction of
the grid parabola,
using arguments from elementary number theory.
\iflong
The final Section\else
Appendix\fi~\ref{experiments}
reports computer experiments with grid peeling for parabolas.
These experiments were the source the discoveries expressed in Theorem~\ref{th:reproduce} below,
and its consequence, our main
Theorem~\ref{main-parabola}.
We also describe some interesting
phenomena beyond
\iflong
those that are discussed and proved in
the first part of the paper.
\else
the ones discussed in
the main part of the paper.
\looseness-1
\fi

\section{The grid parabola}
\label{sec:grid-parabola}
\begin{figure}
    \centering
\includegraphics{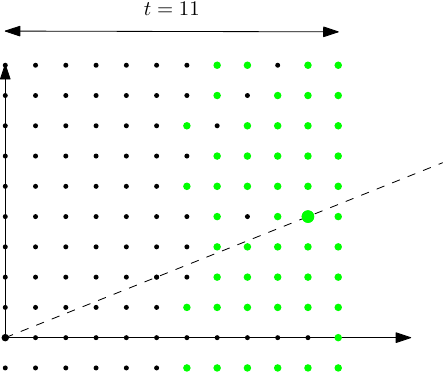}
    \caption{
    The set $V_{11}$ of vectors $(x,y)$ from which $P_{11}$ is formed, shown as green dots.
    The vector with slope $s=2/5$ is highlighted. The points of $V_{11}$ extend indefinitely 
    to the top and to the bottom.
    \iflong The picture shows 
    the range $-1\le y\le9$.\fi
    }
    \label{fig:S11}
\end{figure}

Our object of investigation is a special infinite polygonal chain $P_t$, which depends on
a positive integer parameter~$t$. It is defined as follows:
\begin{enumerate}
\item
Let $S_t$ be the set of all rational numbers
$s=a/b$ with $0< b\le t$.
We call these elements the
\emph{slopes}.
We will always assume that the fractions $a/b$ representing slopes 
are reduced.

\item 
For each slope $s=a/b\in S_t$,
take the longest integer vector of the form 
\begin{displaymath}
\textstyle\binom xy = q\binom{b}{a}
\quad ( q \in\mathbb Z)
\end{displaymath}
with $0< x\le t$.
Let $V_t$ denote the set of these vectors.
Figure \ref{fig:S11} shows 
$V_{t}$ 
for $t=11$.
    \item 
    Form the chain $P = P_t$ by concatenating these vectors in order of increasing slope.
\end{enumerate}

\begin{figure}[htb]
    \centering
    \iflong
    \includegraphics[scale=0.9]{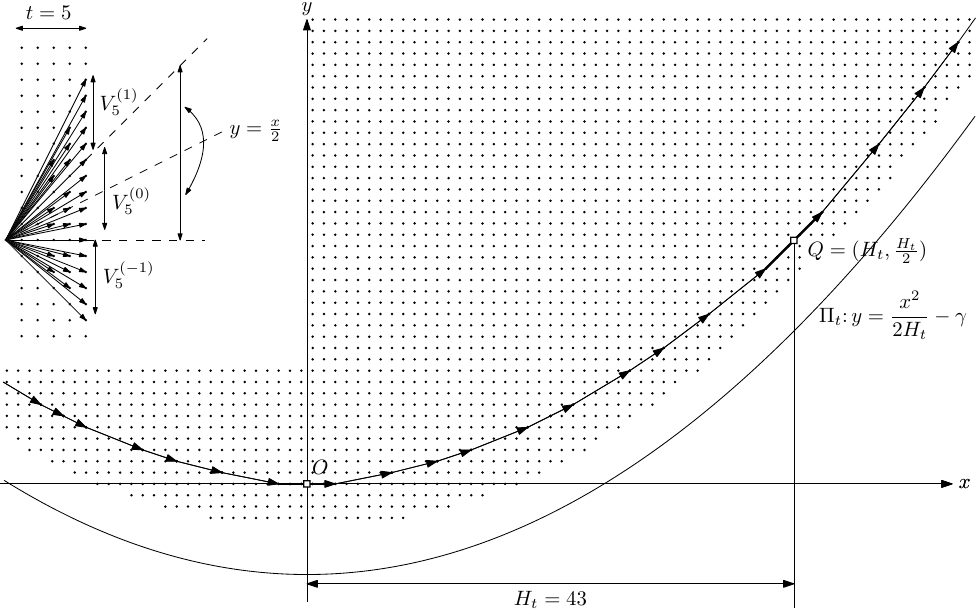}
    \else
    \hskip -3mm\includegraphics[scale=0.85]{P5.pdf}
    \fi
    \caption{The grid parabola $P_t$ for $t=5$. 
    The inset in the upper left corner shows some vectors of the set~$V_5$, at a slightly enlarged scale.
    $\Pi_t$ is the reference parabola defined by $y= \frac{x^2}{2H_t}$.
    Here it is shifted down by some offset~$\gamma$.}
    \label{fig:P5}
\end{figure}

Figure \ref{fig:P5} shows a section of the grid parabola $P_5$.

We can make a few simple observations: 
It is clear that for every vector $(x,y)\in V_t$ with a positive slope, there is a corresponding vector
 $(x,-y)\in V_t$ with negative slope.
Thus, the curve $P$
is symmetric with respect to a vertical axis.  The lowest points on
$P$ form a horizontal edge of length $t$.  We place the origin $O$ of
our coordinate system at the center of this edge, so that the symmetry axis becomes the $y$-axis.
When $t$ is odd,
this implies that the vertices of $P$ have half-integral
$x$-coordinates.
%
%
Nevertheless, we will refer to the points of the
square unit grid on which the vertices of $P$ lie (as shown in Figure \ref{fig:P5}) as
 the \emph{grid
 }.

\begin{figure}
    \centering
    \includegraphics[scale=0.8]{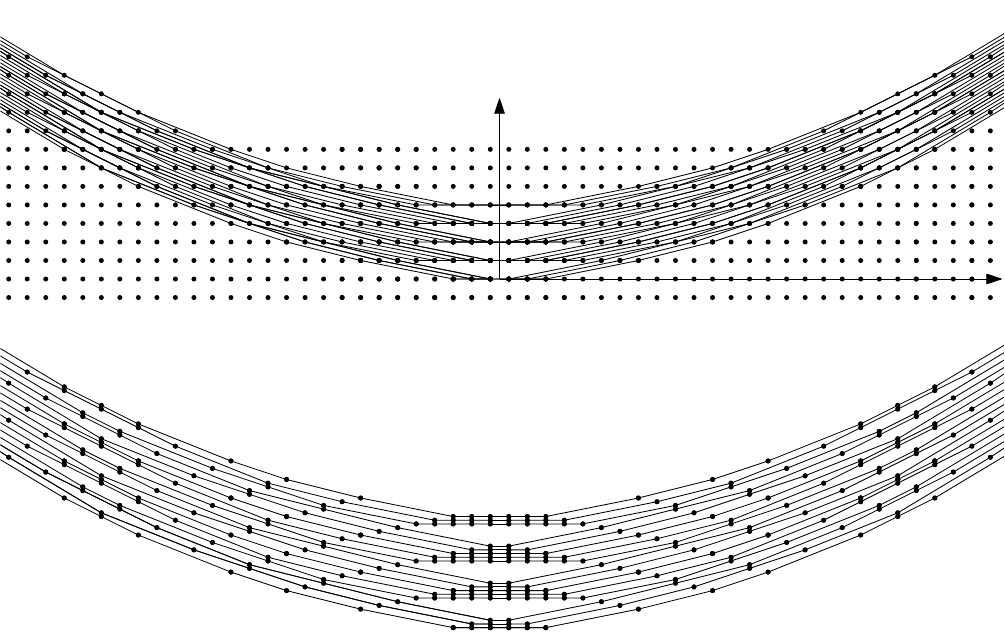}
    \caption{Consecutive peelings of $P_5$.
    Since consecutive peelings share many vertices, it is not easy to
    distinguish the 
    curves.
    In the lower part, we have therefore vertically separated
    the consecutive peelings.
    This has the effect that some grid points appear in several copies with small vertical offsets, and horizontal grid lines get a curved appearance.}
    \label{fig:peel5}
\end{figure}

Figure~\ref{fig:peel5}
applies a few grid peeling steps 
to the grid parabola $P_5$.
We can see that $P_5$ reproduces itself after 5 iterations, translated up by 1 unit.
From then on,
the process repeats 
ad infinitum.
Our central technical result is that this is always the case.

\begin{theorem}\label{th:reproduce}
For odd $t$, the chain $P_t$ repeats after $t$ peeling steps, one unit higher.

For even $t$, the chain $P_t$ repeats after $t+1$ peeling steps, one unit higher.
\end{theorem}

\iflong
Theorem~\ref{th:reproduce} and the special construction of the grid parabola were suggested by experiments, which are reported in Appendix~\ref{experiments}.
In Appendix~\ref{min-area} we mention that our grid parabolas play a role in the convex lattice $n$-gons of minimum area.
\fi

\subsection{The horizontal period}

While $P$ is an infinite object, we will argue that it is
sufficient to look at a finite section, because this section ``repeats periodically'' in a certain sense.

We partition the vectors $V_t = \cdots\cup V_t^{(-2)}\cup V_t^{(-1)}\cup
V_t^{(0)}\cup V_t^{(1)}\cup V_t^{(2)}\cup\cdots$
according to the integral part of their slope into the sets
\begin{displaymath}
V_t^{(i)} := \{\,(x,y)\in S_t \mid i<\tfrac yx \le i+1\,\}
\end{displaymath}
for $i\in \mathbb Z$. The vectors in $V_t^{(0)}$ lead $P$ from the origin to an edge with vector
$(t,t)$ of slope 1.  More precisely, in the way we have defined $V_t^{(0)}$,
these vectors lead from the right endpoint of the horizontal edge to
the upper-right endpoint of the edge $(t,t)$. However, we prefer to
select the midpoint of the edge~$(t,t)$, and
we place a reference point $Q$ at this point.
We
define the \emph{horizontal period $H_t$} as the horizontal distance
between the origin~$O$ and~$Q$.

$H_t$ is the sum of the $x$-coordinates of the vectors in $V_t^{(0)}$.
(Only half of the vector $(t,t)\in V_t^{(0)}$ contributes to $H_t$, but
this is compensated by including
half of the vector $(t,0)\notin V_t^{(0)}$.)
The first values of $H_t$ are
$H_1,H_2,\ldots = 1,4,11,22,43,64,107,150,
$ etc.
In Section~\ref{Ht}, we will evaluate this quantity and see that
 $H_t \approx 0.24 \, t^3$ (Lemma~\ref{lem-Ht-asymptotic}).

We claim that the segment $OQ$ has slope $\frac12$, and therefore
the $y$-coordinate of $Q$ is $H_t/2$.
\begin{proposition}\label{prop:parity}
\iflong
\begin{enumerate}
    \item 
    \label{slope-half}
\fi    
    The segment $OQ$ has slope $\frac12$.
\iflong    
\item
$H_t$ is even if and only if $t$ is even.
    \label{parity}
\end{enumerate}
\fi
\end{proposition}
\begin{proof}
The set $V_t$ is symmetric with respect to the
shearing operation
$\binom xy\leftrightarrow \binom x{x-y}$, which keeps the ``mirror line'' $y=x/2$
fixed and inverts the orientation of every vertical line,
see the inset of Figure~\ref{fig:P5}.
 Thus, every vector in $V_t$ with slope $>\frac12$
can be matched with a vector with slope $<\frac12$, so that the sum of
these two vectors has slope $\frac12$.
The set $V_t^{(0)}$ is slightly asymmetric with respect to this mirror
operation because it contains
the edge of slope~1
but not the corresponding edge of slope~0.
This asymmetry is taken care of by including half of both edges in the
vector from $O$ to $Q$.
Thus, the slope between $O$ and $Q$ averages
\iflong out  \fi 
to $\frac12$.
\iflong

To see statement~\ref{parity}, note that the vectors that are matched in a pair sum to a vector with an even $x$-coordinate. The unmatched vectors in 
 $V_t^{(0)}$ are the vector of slope $s=\frac12$, which has an even $x$-coordinate,
 and the vector $\binom tt$, whose parity therefore decides the
 parity of~$H_t$.
\fi 
 \iflong\else\looseness-1\fi
\end{proof}

\subsubsection{Periodic continuation}
The mapping $(x,y)\mapsto (x,y+ix)$ maps $V_t^{(0)}$ to $V_t^{(i)}$; hence it
is sufficient to know $V_t^{(0)}$; all other sets $V_t^{(i)}$ are copies of $V_t^{(0)}$
where the slope of each vector is modified by an integer constant,
leaving the $x$-coordinate fixed.
This means that the continuation of $P$ beyond the arc from $O$ to $Q$
is in some sense periodic: the same sequence of edges will appear
again and again, only with modified slopes.
The mapping
$(x,y)\mapsto (x+H_t,y+H_t/2+x)$
maps $O$ to~$Q$, and it
maps the curve $P$ to itself. The midpoints of the edges with integer slopes appear regularly at
intervals of length $H_t$. The midpoint of the
edge with slope $i$ is $(iH_t,i^2H_t/2)$.
These points lie on a parabola, which we call
the \emph{reference parabola}
\begin{displaymath}
    \Pi_t \colon y=x^2/(2H_t),
\end{displaymath}
and the polygonal chain $P_t$ follows $\Pi_t$ with bounded local deviations. We summarize
these considerations in the following lemma,
whose proof is straightforward.
\begin{lemma}\label{lem:period-Pt}
    The affine transformation
    \begin{displaymath}
    \binom{x}{y}
    \mapsto
    \binom{x+H_t}{y+x+H_t/2}
    \end{displaymath}
    maps the grid to itself, and in addition, it maps both
 $P_t$ and the reference parabola $\Pi_t$, or any vertical translate of it, to itself.
 \qed
\end{lemma}

\section{Grid peeling for parabolas}


We start with the simple observation that grid peeling preserves inclusion:
 \begin{observation}
   \label{monotonicity}
   Let $U\subseteq U'\subset \mathbb Z^2$ be two sets of grid points that are \emph{upward closed}: $(x,y)\in U\implies (x,y+1)\in U$. Let $f$ denote one peeling step.
   Then $f(U)\subseteq f(U')$. \qed
 \end{observation}

Observation~\ref{monotonicity} implies that if $C$ and $D$ are two convex $x$-monotone curves that extend from $x=-\infty$ to $x=+\infty$ (e.g. grid curve or an arbitrary smooth or piecewise smooth curve), and $C$ lies everywhere (weakly) below $D$, then this relation is maintained by grid peeling.

\label{sec:main-result}
 \begin{lemma}
   $P$ and $\Pi$ advance at the same limiting speed.
 \end{lemma}

 \begin{proof}
   As already argued, $P$ approximates $\Pi$ in a global
   sense, while locally, there might be deviations.
   This implies that
   we can shift $\Pi$ vertically down by some integer amount~$\gamma$ and ensure that the shifted parabola, denoted by $\Pi-\gamma$, lies completely
   below $P$.
Figure~\ref{fig:P5} shows the parabola for $\gamma=8$, but actually, $\gamma=1$
should already be sufficient to
push $\Pi$ below $P$.

Now imagine that we start grid peeling simultaneously with $P$ and with
$\Pi-\gamma$, or more precisely, with the convex hull of the grid points on or
above $\Pi-\gamma$.   
Applying the {monotonicity} property of Observation~\ref{monotonicity},
we conclude 
that the evolution of
$\Pi-\gamma$ always remains below the evolution of $P$.
It can never overtake $P$, and in particular, the limiting speed of
$\Pi-\gamma$, which is the same as the limiting speed of $\Pi$, is at most
the limiting speed of~$P$.

We can push $\Pi$ upward and start with a parabola $\Pi+\gamma'$ that lies
above $P$ everywhere, and argue in the same way that
the evolution of $P$ can never overtake the evolution of $\Pi+\gamma'$, and
thus,
the limiting speed of $\Pi$ is at least
the limiting speed of $P$.
 \end{proof}

\subsection{Evaluating the horizontal period \texorpdfstring{$H_t$}{H\_t}}
\label{Ht}

We have seen that $H_t$ is the sum of the $x$-coordinates of the vectors in $V_t^{(0)}$.
It 
is thus given by the following expression:
  \begin{equation}\nonumber %
    H_t = \sum_{(x,y)\in V_t^{(0)}} x
=
    \sum_{\substack{  0<y\le x\le t\\\gcd(x,y)=1}}
  \left\lfloor \frac
    tx \right\rfloor x 
  \end{equation}
This sequence appears in
\iflong
the Online Encyclopedia of Integer Sequences
\else
the O.E.I.S.\ 
\fi
\cite[\href{https://oeis.org/A174405}{A174405}]{OEIS}.
It starts with the values
\begin{displaymath}
  H_1,H_2,\ldots =
  1, 4, 11, 22, 43, 64, 107, 150, 211, 274, 385, 462, 619, 748, 895, 1066, 1339, 
\ldots
\end{displaymath}
\iflong
The sequence 
has been investigated by Sándor and Kramer
\cite{sandor99},
who showed the following 
asymptotic estimate: 
\begin{lemma}\label{lem-Ht-asymptotic-weak}
\begin{displaymath}
    H_t     \sim 
 \frac {2\zeta(3)}{\pi^2}
 t^3 
\approx 0.243587656\, t^3
\end{displaymath}
\end{lemma}\else
Sándor and Kramer
\cite{sandor99}
proved the
asymptotic estimate $H_t
\sim 
 \frac {2\zeta(3)}{\pi^2}
 t^3 \approx 0.243587656\, t^3
$.
\fi
This can also be derived as 
a consequence of the more general Lemma~\ref{vectors-in-triangle}, which is stated 
below.
Section~\ref{Ht-alternative}
  gives several alternative expressions for $H_t$.

\subsection{Distance between the grid parabola and the reference parabola}

We want to analyze the deviation between the grid parabola and the reference parabola. While the reference parabola has an explicit expression, 
the grid parabola is given by a multistep process, as described in 
Section~\ref{sec:grid-parabola}.
For getting 
from the origin to an arbitrary vertex of $P_t$,
we need to sum the vectors whose slope is at most some threshold $\alpha$:
\begin{displaymath}
     U_t^\alpha
     :=
    \sum_{\substack{  0\le x\le t\\0<y\le \alpha x\\\gcd(x,y)=1}}
  \left\lfloor \frac
    tx \right\rfloor \binom xy 
\end{displaymath}

More precisely, since
the sum includes only vectors with $y>0$,
it measures the distance from the right endpoint of the 
horizontal segment of $P_t$ and not from the origin.
Lemma~\ref{vectors-in-triangle},
whose proof will be given in Appendix~\ref{proof-alpha},
gives an
asymptotic expression for this sum:

\begin{lemma}\label{vectors-in-triangle}
\label{lem-Ht-asymptotic}
    Let $0\le\alpha\le 1$.
Then
      \begin{displaymath}
      U_t^\alpha 
= 
 \frac {2\zeta(3)}{\pi^2} 
 \binom
 {t^3\alpha  + O(t^2\log t)}
 {t^3\alpha^2 /2 + O(t^2\log t)},
 \text{ \ and  \ }
 H_t  =
    \frac {2\zeta(3)}{\pi^2}
    t^3 + O(t^2\log t).
      \end{displaymath}
\end{lemma}
The second expression is obtained from
the first one by setting $\alpha=1$ and looking only at the $x$-coordinate.

\begin{proposition}\label{offset-reference}
The vertical distance between the grid parabola $P_t$ and the reference parabola
$\Pi_t$ is bounded by $O(t^2\log t)$.
\end{proposition}

\begin{figure}
    \centering
    \includegraphics{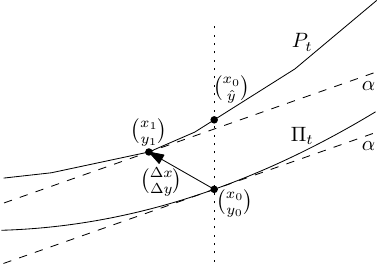}
    \caption{The vertical distance between $P_t$ and $\Pi_t$}
    \label{fig:vertical-distance}
\end{figure}

\begin{proof}
By the periodic behavior of $P_t$ and $\Pi_t$ (Lemma~\ref{lem:period-Pt}),
it suffices to look at the interval 
$0\le x \le H_t$.
Pick a point $x_0$ in this interval, see Figure~\ref{fig:vertical-distance}. The corresponding point
 $\binom{x_0}{y_0}$ on $\Pi_t$
has $y_0=x_0^2/(2H_t)$, and the slope at this point is $\alpha := x_0/H_t\le 1$.
As a first step,
we find the point $\binom{x_1}{y_1}$ on~$P_t$ with the same slope~$\alpha$. We will show that it deviates from 
$\binom {x_0} {y_0 }$ by at most $O(t^2\log t)$ in each coordinate:
By construction
the grid parabola contains the vertex
\begin{displaymath}
    \binom{x_1}{y_1} =\binom{t/2}{0} +
 U_t^\alpha .
\end{displaymath}
The correction term $t/2$ 
accounts for the fact that
$U_t^\alpha$ does not include the vector $\binom t0$ and thus
measures the distance from the right endpoint of the 
horizontal segment of $P_t$ and not from the origin.
Applying both parts of Lemma~\ref{vectors-in-triangle},
we get
\iflong
\begin{align*}
\binom{x_1}{y_1} =\binom{t/2}{0} +
 U_t^\alpha 
  & =\frac{2\zeta(3)t^3}{\pi^2}
    \binom
 {\alpha  }
 {\alpha^2 /2 }
  + \binom{O(t^2\log t)}{O(t^2\log t)} + \binom{t/2}{0}
   \\ & \nonumber
=\bigl(H_t+O(t^2\log t)\bigr)
 \binom
 {\alpha  }
 {\alpha^2 /2 }
+
   \binom{O(t^2\log t)}{O(t^2\log t)}
   \\ & \nonumber
=H_t
 \binom
 {\alpha  }
 {\alpha^2 /2 }
+
   \binom{O(t^2\log t)}{O(t^2\log t)}  
   \\ & \nonumber
=
 \binom
 {x_0}
 {y_0 }
+
    \binom{\Delta x}{\Delta y}
\end{align*}
\else

\begin{align*} 
\binom{x_1}{y_1} =\binom{t/2}{0} +
 U_t^\alpha 
  & =\frac{2\zeta(3)t^3}{\pi^2}
    \binom
 {\alpha  }
 {\alpha^2 /2 }
  + \binom{O(t^2\log t)}{O(t^2\log t)} + \binom{t/2}{0}
   \\ & \nonumber
=\bigl(H_t+O(t^2\log t)\bigr)
 \binom
 {\alpha  }
 {\alpha^2 /2 }
+
   \binom{O(t^2\log t)}{O(t^2\log t)}
   \\ & \nonumber
=H_t
 \binom
 {\alpha  }
 {\alpha^2 /2 }
+
   \binom{O(t^2\log t)}{O(t^2\log t)}  
=
 \binom
 {x_0}
 {y_0 }
+
    \binom{\Delta x}{\Delta y}
\end{align*}
\fi
with
$\Delta x,\Delta y =O(t^2\log t)$. In the range $0\le x \le H_t$,
the slope of $P_t$ is bounded by 1. So when we move from $x_1$ to $x_0=x_1+\Delta x$ on $P_t$,
we arrive at a point
$(x_0,\hat y)$ with $|\hat y-y_1|\le |\Delta x|$,
and thus, the vertical distance $|\hat y-y_0|$ between $P_t$ and $\Pi_t$
is at most $|\Delta y|+|\Delta x|=O(t^2\log t)$.
\end{proof}

Figure~\ref{fig:diff-reference} in Appendix~\ref{appendix-difference} shows the actual difference $P_t-\Pi_t$ for a few selected values of $t$.

\subsection{Comparison to true parabolas}


Let $y=ax^2/2 + bx + c$. We are interested in the average (vertical) speed in which
the curve moves upwards. 
If we start grid peeling with a parabola
 $y=ax^2 + bx + c$ for rational coefficients $a$ and $b$, we can show that, after some irregular ``preperiod'', it
 will enter a periodic behavior:
 After a certain number $\Delta m$ of steps, the same curve reappears, translated upward by $\Delta y$. We call
$\Delta m$ the \emph{time period} and
$\Delta y$ the \emph{vertical period}
(to be distinguished from yet another period, the horizontal period~$H$,
which was introduced at the beginning).
The \emph{average (vertical) speed} is then 
$\Delta y/\Delta m$. 
However, if $a$ or $b$ is irrational, we no longer have a periodic
behavior.
For a more general curve, like $y=e^x$, different parts of the curve will move at different speeds, and a common average speed will not exist. 
For this reason, we define the \emph{lower} and \emph{upper} average speed.

\begin{definition}[average speed]
    Let $C$ be the graph of a convex function on $\mathbb R$. We denote by $C+\gamma$ the copy of $C$ vertically translated by
   $\gamma$ (upwards for $\gamma>0$, downwards for $\gamma<0$).

   For another such curve $D$, 
   we write $C\le D
   $ if no point of
   $C$ lies above~$D
   $.

   Let $f^{(m)}$ denote $m$ steps of grid peeling.

The \emph{lower average speed} $v_- = v_-(C)$ is defined as follows:
\begin{displaymath}
    v_- = \liminf_{m\to\infty} \frac
    {\sup \{\,\gamma\mid C+\gamma \le f^{(m)}(C)\,\}}{m}
\end{displaymath}
The \emph{upper average speed} $v_+ = v_+(C)$ is defined similarly:
\begin{displaymath}
    v_+ = \limsup_{m\to\infty} \frac
    {\inf \{\,\gamma\mid f^{(m)}(C)\le C+\gamma \,\}}{m}
\end{displaymath}
If $v_-$ and $v_+$ coincide, we call it simply the
 average speed $v=v(C)$. 
\end{definition}

\begin{figure}
    \centering
    \includegraphics{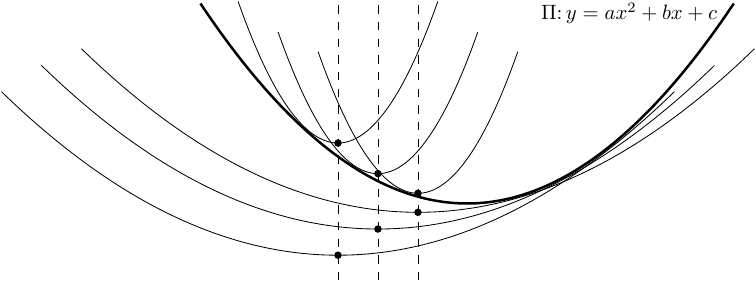}
    \caption{Upper and lower approximation
    of the parabola $\Pi$ by ``integer'' parabolas}
    \label{fig:inner-and-outer-parabolas}
\end{figure}

\iflong
We have the obvious inequalities
$0\le v_-\le v_+\le\infty$.
\fi
By approximating the parabola
 $y=ax^2/2 + bx + c$ from above and below by appropriate
 grid parabolas (see
Figure~\ref{fig:inner-and-outer-parabolas}),
to which we apply Theorem~\ref{th:reproduce},
we arrive at the following result:

\goodbreak

\begin{theorem} \label{average-speed}
    \begin{enumerate}
        \item 
        If $\frac1{H_{t}} < a < \frac1{H_{t-1}}$
        and $t$ is odd
        \textup(or $a>\frac1{H_{1}}=1$\textup), then $v=1/t$.
        \item 
        If $\frac1{H_{t}} < a < \frac1{H_{t-1}}$
        and $t$ is even, then $\frac 1{t+1}\le v_-\le v_+ \le \frac1{t-1}$.
        \item 
        If $a = \frac1{H_{t}}$ and $t$ is odd, then $\frac 1{t+2}\le v_-\le v_+ \le \frac1{t}$.
        \item 
        If $a = \frac1{H_{t}}$ and $t$ is even, then $\frac 1{t+1}\le v_-\le v_+ \le \frac1{t-1}$.
        \qed
    \end{enumerate}
\end{theorem}
The experiments in Section~\ref{experiments} suggest that the first statement also holds for even $t$, and for
 $a = \frac1{2H_{t}}$,
$v$ exists always and lies in the range 
 $\frac 1{t+1}\le v \le \frac1{t}$.

\begin{proposition}\label{invariant-translations}
Let $C$ be the graph of a convex function on $\mathbb R$.
    The upper and lower average 
    speed is not changed by
    \begin{itemize}
        \item horizontal translation by an integer distance,
        \item or arbitrary vertical translation.
    \end{itemize}
\end{proposition}
\begin{proof}
    It is clear that a translation by an integer vector does not change anything.
    
    Consider an arbitrary vertical translation $C+\gamma$.
    Then (1)~The integer translates $C_0=C+\lfloor\gamma\rfloor$ and
   $C_1=C+\lfloor\gamma\rfloor+1$ have the same 
   upper and lower average 
   speeds as~$C$,
   (2)~they maintain a constant vertical distance of 1 during grid peeling,
   and (3)~$C+\gamma$ is sandwiched between $C_0$ and $C_1$, and it maintains this relation during grid peeling.
   It follows that
   the upper and lower average 
   speeds of $C$ must agree with those of $C_1$ and $C_2$.
\end{proof}

One might be tempted to believe that a small horizontal translation should also
not change the vertical speed. However,
Section~\ref{b-at-critical} reports examples where translations cause the
 vertical speed to change, see
Figure~\ref{fig:critical}.
 
\subsection{Refined grid peeling for parabolas, proof of Theorem~\ref{main-parabola}}

We prove
our main theorem,
Theorem~\ref{main-parabola}
about the relation between grid peeling and ACSF
for parabolas
$y=ax^2/2 + bx + c$.
We use $(x,y)$ for the original coordinates, with a grid of spacing $1/n$, and
$(\hat x,\hat y)=(nx,xy)$ for the scaled coordinates, with a unit grid. The curvature at the vertex of the parabola is $a$; thus the vertical speed of ACSF at this point (and thus everywhere, by
affine invariance) is $a^{1/3}$.


At time $T$ we have $y=ax^2/2 + bx + c+ Ta^{1/3}$, and $\hat y=\frac an \hat x^2/2 + b\hat x + cn + Tna^{1/3}$. Determine $t$ such that $\frac{1}{H_t} \le \frac an \le \frac{1}{H_{t-1}}$. So $\frac an \approx \frac{1}{H_t}$ and Lemma~\ref{lem-Ht-asymptotic} implies
\begin{displaymath}
\frac na \approx H_t =
\frac{2\zeta(3)  t^3}{\pi^2} 
\cdot (1+ \textstyle O(\log t/t))
=\Bigl(\dfrac{t}{c_{\mathrm{g}}}\Bigr)^3
\cdot (1+ \textstyle O(\log t/t)),
\end{displaymath}
which yields
\begin{displaymath}
t = c_{\mathrm{g}}\sqrt[3]{\frac n{a\bigl(1+ \textstyle O(\log t/t)\bigr)}} +O(1) =
c_{\mathrm{g}}\sqrt[3]{\tfrac na}\cdot \bigl(1+  O(\log t/t)\bigl)
=\Theta\Bigl(\sqrt[3]{\tfrac na}
\Bigr)
.
\end{displaymath}
Here, the 
$O(1)$ term accounts for rounding $t$ to an integer,
and it also covers the
 uncertainty of  Theorem~\ref{average-speed}, where
$t$ is sometimes replaced by $t-1$, $t+1$, or $t+2$.
This additive error term is absorbed in the multiplicative error
term $1+ O(\log t/t)$.
By Theorem~\ref{average-speed}, the lower and upper average speed is
\begin{displaymath}
    v\approx \frac 1{t
    } = 
    \frac 1{c_{\mathrm{g}}}\sqrt[3]{\tfrac an}\cdot 
        \bigl(1+ \textstyle O(\log t/t)\bigl)
\end{displaymath}
After
$m = \lfloor c_{\mathrm{g}} T n^{4/3} \rfloor$
steps, the 
vertical distance that the curve has moved up is therefore
\begin{displaymath}
    mv +O(1) =  Ta^{1/3}n\bigl(1+  O(\log t/t)\bigl) +O(1)
    =
    Ta^{1/3}n\bigl(1+ O(\sqrt[3]{\tfrac an}\log \tfrac na)\bigl)
\end{displaymath}
The difference to the movement of $\Pi$, which is $Tna^{1/3}$,
is
\begin{displaymath}
O\bigl(Ta^{1/3}n\sqrt[3]{\tfrac an}\log \tfrac na\bigr)
=
O(T
{a^{2/3}}{n^{2/3}}\log \tfrac na).
\end{displaymath}
To this, we must add the distance between $P_t$ and the reference parabola $\Pi_t$
from Proposition~\ref{offset-reference}, that is,
$O(t^2\log t) = O((\frac na)^{2/3} \log \tfrac na)$.
Dividing by $n$, we conclude that the error term in terms of the original $y$-coordinates is
\begin{displaymath}
\bigl(O(T{a^{2/3}}{n^{2/3}}\log \tfrac na)
+O((\tfrac na)^{2/3} \log \tfrac na) 
\bigr)/n
=
O\bigl((Ta^{2/3}+a^{-2/3})/n^{1/3} \log \tfrac na\bigr). \hfill\qed
\end{displaymath}

\section{Proof of Theorem~\ref{th:reproduce} about the period of 
the grid parabola}
\label{sec:proof-grid}

When we speak of \emph{the curve}, we mean the grid parabola $P_t$ after some iterations of peeling.

\begin{figure}[htbp]
    \centering
    \includegraphics[scale=0.9]{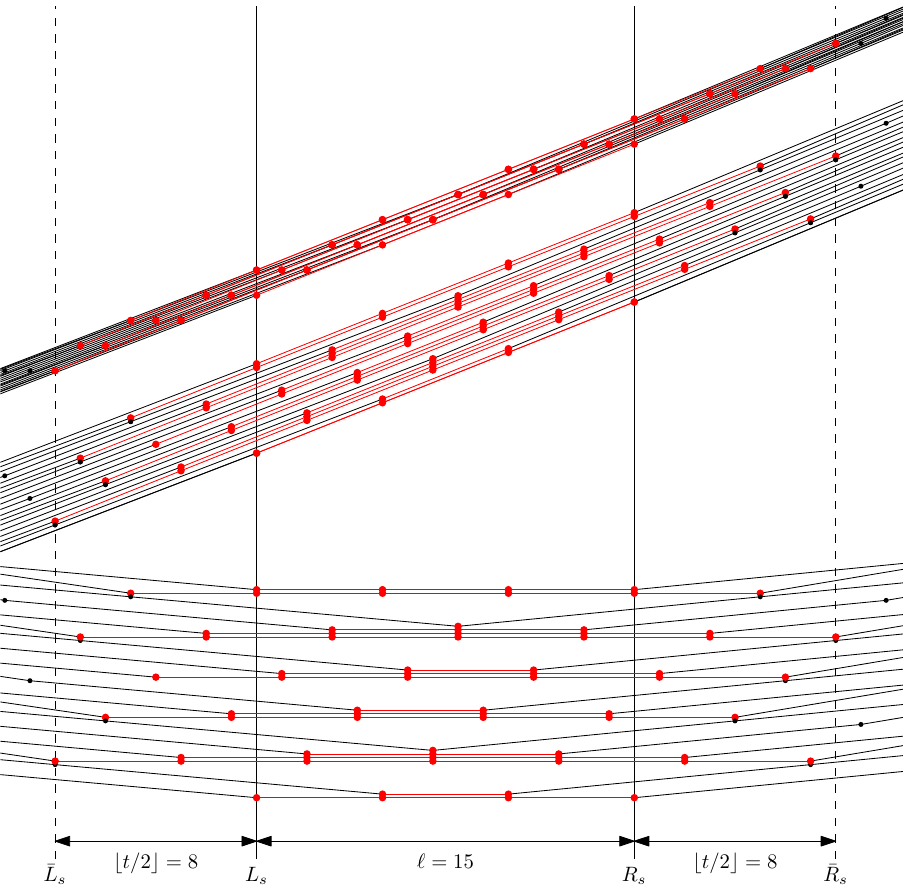}
    \caption{17 \iflong consecutive \fi iterations of the development of slope $s=2/5$ for $t=16$, starting with the curve~$P_t$.
    The uppermost part shows the true situation.
    In the middle part, the successive curves are separated for better visibility, as in Figure~\ref{fig:peel5}.
    In the lowest part, an affine transformation has been applied to make
    the segment of interest horizontal; this allows the different slopes
    to be distinguished more easily.
    The segment of slope $s$ is highlighted in red on each curve.
    %
    The initial segment on $P_{16}$ has horizontal length $\ell
    =\lfloor\frac{16}{5}\rfloor\times5 =15$.
    The region between the dashed lines is the extended strip.
   }
    \label{fig:one-slope}
\end{figure}

Let $s=a/b\in S_t$ be a fixed slope.
We consider the supporting line $g$ with slope $s$, and we study how it evolves during the peeling process, see Figure~\ref{fig:one-slope} for an illustration.

\begin{definition}
The \emph{strip} of slope $s$ is the vertical strip that bounds the
segment of slope $s$ in $P_t$.
It goes from $x=L_s$ to $x=R_s$.
We denote by $\ell = R_s-L_s$ the \emph{width} of the strip.

The \emph{extended strip} of slope $s$ includes an additional margin of 
$\lfloor t/2\rfloor$ on both sides.
It goes from  $\bar L_s=L_s-\lfloor t/2\rfloor$ to $\bar R_s = R_s+\lfloor t/2\rfloor$.
\end{definition}

We state some obvious 
properties of the peeling process:

\begin{observation}\label{simpleproperties}
Throughout the whole peeling process:
    \begin{enumerate}[(i)]
    \item The supporting line $g$ intersects the curve in a 
    \iflong    line segment, which might degenerate to a point.
    \else line segment or a point.
    \fi  
    \label{1}
    \iflong\looseness-1\fi
    \item If the segment contains $k\ge1$ grid points, its horizontal length is $(k-1)b$.
    \item At every peeling step, the two endpoints of this line segment or the single point is peeled.
    \iflong\looseness-1\fi
    \item As long as the segment contains at least 3 grid points, the supporting line does not change, and the number $k$ of grid points on the segment decreases by~2. In this case, we say that the segment \emph{shrinks}.
    \item If the segment contains only 1 or 2 grid points, the supporting line changes. We say that there is a \emph{jump} for slope~$s$.\label{easy} \qed
    \end{enumerate}
\end{observation}

We use the following terminology:
The \emph{left endpoint of slope $s$} is the left endpoint of the segment
of slope $s$; in case the segment degenerates to a single point, it is that point.
In other words, it is the leftmost point where the supporting line of slope $s$ touches the curve.
The \emph{right endpoint} is defined analogously.
We will only deal with \emph{horizontal} offsets, lengths, positions,
and distances, and thus
we will often omit the word 
\emph{horizontal}.

In the following crucial lemma, 
Properties~\ref{next-line} and
~\ref{fill-strip} predict \emph{what} happens when a jump occurs.
In particular, Property~\ref{fill-strip} characterizes
the possible locations of the vertices after a jump.
Property~\ref{jump-position} describes the final position of the segment before the jump.
This statement allows us to predict \emph{when} a jump occurs.
Property~\ref{jump-position} can be easily worked out,
\emph{assuming} that the initial position after the previous jump satisfies Property~\ref{fill-strip}. 
Properties~\ref{endpoints-match} and
~\ref{breakpoint-position} describe the situation when two consecutive slopes are involved. 

\begin{figure}
    \centering
    \includegraphics[scale=0.9]{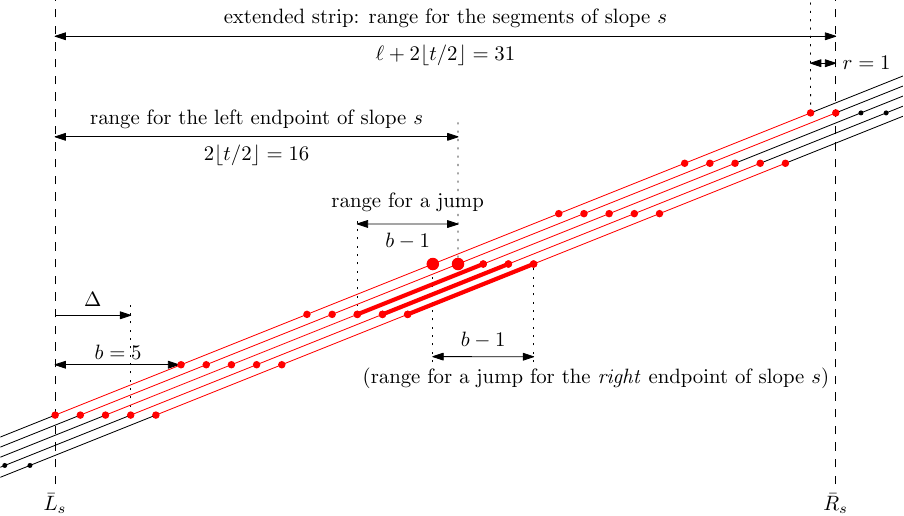}
    \caption{The $b=5$ grid lines
    from Figure~\ref{fig:one-slope}.
    $t=16=3b+1=qb+r$.
    The initial segment on each line, immediately after the jump, is shown in red.
    The lines are considered according to the offset $\Delta$ of the leftmost grid point from the left edge of the extended strip.
    We have highlighted
    the last remaining single point or pair of points before the jump occurs.
    }
    \label{fig:one-slope-ordered}
\end{figure}

\begin{lemma}\label{predict}
The following properties hold throughout the peeling process:
\begin{enumerate}
    \item (No grid line is skipped.) Whenever there is a jump, the supporting line $g$ of slope $s$
    advances to the \emph{next grid line} with slope $s$. \label{next-line}
    \item (The filling property.) After the supporting line $g$ has advanced, the curve will contain precisely those grid points on $g$ that lie within the extended strip of $s$.
    In other words, the segment fills the extended strip as much as possible. (See Figure~\ref{fig:one-slope} and~\ref{fig:one-slope-ordered}.)
    \label{fill-strip}
    \item (Jump position) A jump of slope $s$ occurs 
    \iflong if and only if \else iff \fi the left endpoint of slope $s$
    lies in the range
    \begin{equation}
        \label{range-jump}
        L_s+\lfloor t/2\rfloor-(b-1),\ \ldots,\ L_s+\lfloor t/2\rfloor.
    \end{equation}
    A symmetric property holds for the right endpoint.
    \label{jump-position}
    \item (There are no gaps.)
    For any two consecutive slopes $s=a/b$ and $s'=a'/b'$ from $S_t$, with $s<s'$, the right endpoint of slope $s$ coincides with the left endpoint of slope $s'$. In particular, no edge has an intermediate slope between $s$ and $s'$. 
    This implies that only slopes from the set $S_t$ appear in the curves.
    \label{endpoints-match}
    \item (Breakpoint position) The breakpoint between two consecutive slopes $s,s'$ is in the interval between $X-t/2$ and $X+t/2$, where $X = R_s = L_{s'}$ is the boundary between the corresponding strips. (See Figure~\ref{fig:jumps}.)
    \label{breakpoint-position}    
    
\end{enumerate}

\end{lemma}
Properties~\ref{next-line}--
\ref{jump-position},
which were discovered experimentally, are strong enough to
predict the behavior of the supporting segment of slope $s$ during
the peeling process in a purely local manner, without looking at the whole curve. The proof that this is the evolution that actually takes place
amounts to checking whether these local characterizations
fit together when considering different slopes.
In particular, we will look at two consecutive slopes 
(Properties~\ref{endpoints-match} and
~\ref{breakpoint-position}).
This will involve checking some cases, but with the rigid structure
provided by the strong
properties~\ref{next-line}--
\ref{jump-position},
one cannot really avoid to come up with the proof.

\begin{proof}
We will show inductively that the claimed properties are maintained as invariants for all slopes throughout the peeling process.

We rely on the following properties of two consecutive slopes $s=\frac ab$ and $s'=\frac{a'}{b'}$, which follow from the definition of the slope set $S_t$ (Section~\ref{sec:grid-parabola}):
\begin{enumerate}[(I)]
    \item \label{b+b'} The denominators $b$ and $b'$ are bounded by $b,b'\le t$, and 
    their sum is $b+b'>t$.
    
    (Otherwise, the vector $\binom{b}{a}+\binom{b'}{a'}$ would
    give rise to a slope in $S_t$ between $s$ and $s'$.)
    \item \label{basis} The two vectors
    $\binom{b}{a}$ and
    $\binom{b'}{a'}$ form a lattice basis of the unit grid.
    
    (Otherwise, they would span a parallelogram that contains interior points, and some of these points would lead to vectors
    with an intermediate slope between $s$ and $s'$ 
    in~$S_t$.)
\end{enumerate}

As the basis for the induction, it can be seen without computation that
the ``original'' segment of slope $s$ of $P_t$ falls in the pattern of analysis
leading to Property~\ref{jump-position}:
Indeed, it lies centrally in the extended strip.
Thus, when extending it as much as possible within the extended strip and starting the peeling process, the starting segment will
appear during this process, by symmetry. Also, the endpoints of consecutive segments match on $P_t$ by construction, establishing Property~\ref{endpoints-match} and Property \ref{breakpoint-position} at the beginning.

For the induction step we consider two consecutive slopes $s,s'$ and make sure that no matter if they make a jump or not, 
the properties of Lemma~\ref{predict} hold. Thus we have four cases. In each case we prove properties~\ref{next-line},\ref{fill-strip}, \ref{endpoints-match}, \ref{breakpoint-position}, and then assuming properties~\ref{next-line},\ref{fill-strip}, \ref{endpoints-match}, \ref{breakpoint-position}, we prove Property~\ref{jump-position} all at once.

\textbf{Case 1: $s$ and $s'$ both shrink:}
We show that this is not possible. Assume that $s,s'$ both shrink. Since no jump occurs for $s'$, by Property~\ref{jump-position} the rightmost possible position of the left endpoint of $s'$ is 
$L_{s'} + \lfloor t/2 \rfloor -b'$. Symmetrically, the leftmost possible position of the right endpoint of $s$ is 
$R_{s} - \lfloor t/2 \rfloor +b$, and therefore $R_{s} - \lfloor t/2 \rfloor +b
    \le
    L_{s'} + \lfloor t/2 \rfloor -b'$.
Since $L_{s'}=R_s$, it follows that $ -\lfloor t/2 \rfloor +b \le \lfloor t/2 \rfloor -b'$,
or $b+b'\le 2 \lfloor t/2 \rfloor \le t$, which contradicts~\eqref{b+b'}.

\textbf{Case 2: $s$ jumps and $s'$ shrinks:} 
We claim that the endpoint of the shrunken segment for $s'$
arrives at the next grid line with slope $s$, and its position on this line matches the position for the right endpoint of $s$ predicted by Property~\ref{fill-strip}. The left endpoint of $s'$ moves by the vector $\binom{b'}{a'}$, and 
since $\binom{b'}{a'}$ and $\binom{b}{a}$
form a lattice basis (\ref{basis}), the supporting line of slope $s$ will indeed jump
to the next grid line of slope $s$, establishing 
Property \ref{next-line} in this case.

We claim that this new left endpoint of $s'$ is indeed the rightmost grid point on this line in the extended strip for $s$ (establishing Property~\ref{fill-strip}). 
We show that it lies in the extended strip of $s$, but the point \emph{after} this point on the grid line of slope $s$
is already outside the extended strip.

To show the former, consider
the rightmost possible position for the left endpoint of $s'$ before shrinking. It
is $ L_{s'}+\lfloor t/2\rfloor-b'$ by Property~\ref{jump-position}. After shrinking, it is
$ L_{s'}+\lfloor t/2\rfloor=
 R_{s}+\lfloor t/2\rfloor 
 = \bar R_s$, and thus it lies in the extended strip of $s$.
 To show the latter, consider the leftmost possible position for the left endpoint of $s'$ before shrinking. It
is $\bar L_{s'}=X-\lfloor t/2\rfloor$ by Property~\ref{jump-position}. After shrinking, it is $X-\lfloor t/2\rfloor+b'$. The point {after} this point is at offset~$b$, and
$X-\lfloor t/2\rfloor+b'+b >X-\lfloor t/2\rfloor+t\ge
X+\lfloor t/2\rfloor = \bar R_s$, by \eqref{b+b'}, and thus this point is already outside the extended strip.

Therefore, the new left endpoint of $s'$ coincides with the new right endpoint of $s$, establishing Property~\ref{endpoints-match} and Property~\ref{breakpoint-position} in this case.

\textbf{Case 3: $s$ shrinks and $s'$ jumps:}
The situation is symmetric to Case 2.

\begin{figure}
    \centering
    \iflong
      \includegraphics{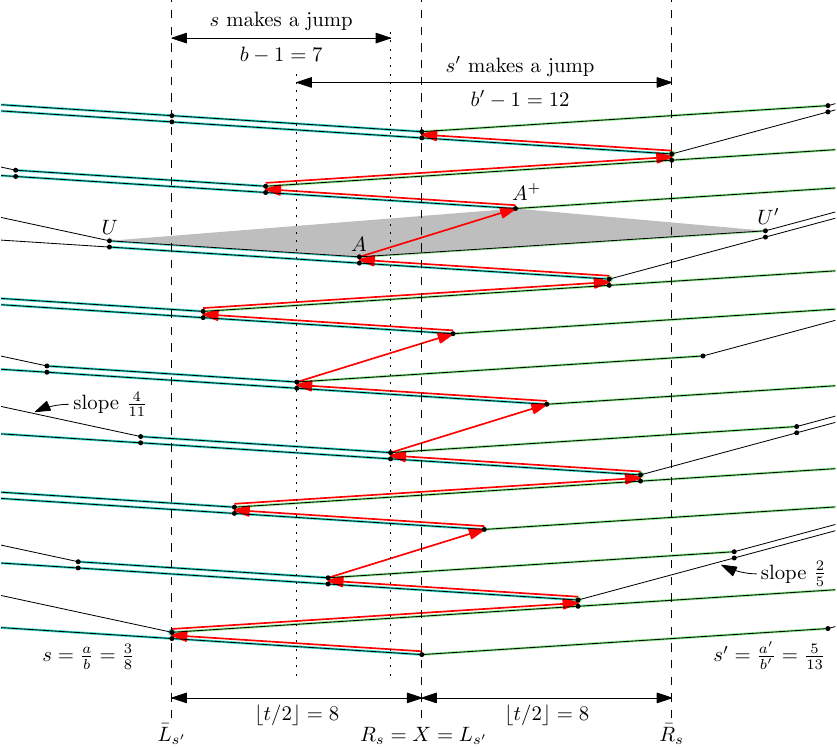}
    \else
      \includegraphics[scale=0.98]{jumps.pdf}
    \fi
    \caption{The transition between slope $s=3/8$, with vector
    $\binom{16}{6}\in V_t$,
    and $s'=5/13$, with vector $\binom{13}{5}\in V_t$, for $t=16$.
    The figure is drawn after an affine transformation, following
    the conventions of the lower part of Figure~\ref{fig:one-slope}.
    The right endpoint of slope $s$ always coincides with the left endpoint of slope $s'$, and it varies in the interval between $X-\lfloor t/2\rfloor$ and $X+\lfloor t/2\rfloor$.}
    \label{fig:jumps}
\end{figure}

\textbf{Case 4: $s$ and $s'$ both jump:}
By Property~\ref{jump-position} the position $x$ of the breakpoint $A=\binom xy$ before the jump is
       in the interval 
\begin{equation}
    \label{range-doublejump}
       X+\lfloor t/2\rfloor-b' < x < X-\lfloor t/2\rfloor+b
\end{equation}
 Let $U$ be the previous point on the grid line of slope $s$ through~$A$,
 and let $U'$ be the next point on the grid line of slope $s'$ through~$A$.
 Construct the parallelogram $UAU'A^+$. 
 Such a parallelogram is shaded in Figure~\ref{fig:jumps}. We know that both supporting lines of slope $s$ and $s'$ must advance at least 
  to the next grid line. Those two grid lines intersect
  at the fourth point $A^+=\binom{x^+}{y^+}$
  of this parallelogram with $x^+= 
  x-b+b'$. We will show two things:
  
  \begin{enumerate}
      \item [(a)]
      The point $A^+$ is not peeled until after the jump.
  Thus it will be the common right endpoint of slope $s$
  and left endpoint of slope $s'$.

  \item [(b)]
  It is indeed the rightmost grid point on the grid line of slope $s$ in the extended strip of~$s$.
  
  (Symmetrically, it 
  is the leftmost point on the grid line in the extended strip of $s'$.)
  \end{enumerate}
\goodbreak
  To prove (a), assume w.l.o.g.\ that
  $b\le b'$, so that
  the segment $AU'$ lies below  $A^+$.
  We are done if we show that
  this segment is part of the boundary when $A$ is peeled.
  Assume otherwise.
  Then $A$ would be not only the left endpoint of slope $s'$, but also the right endpoint of slope $s'$ when it is peeled. According to Property~\ref{jump-position},
  this means that $x \ge R_{s'} - \lfloor t/2 \rfloor$.
  Let $\ell_{s'}\ge b'$ denote the length of the strip of slope~$s'$.
Then, with $b \le b'\le \ell_{s'}$
  we get a contradiction to \eqref{range-doublejump}:
\begin{displaymath}
   x \ge R_{s'} - \lfloor t/2 \rfloor \ge
   R_{s'} -\ell_{s'}  -\lfloor t/2\rfloor+ b
   = X-\lfloor t/2\rfloor+b > x
\end{displaymath}

  To prove (b), note that
  $A^+$ lies in the extended strip for $s$ because
  $x^+=x-b+b'< X - \lfloor t/2\rfloor+b'\le
  X - \lfloor t/2\rfloor+t$, by
  the right inequality of \eqref{range-doublejump},
  and hence
  $x^+\le X + \lfloor t/2\rfloor$. On the other hand, the next grid point on the line of slope $s$
  has $x$-coordinate $x^+ +b= x+b'>X+\lfloor t/2\rfloor$, by the left inequality of \eqref{range-doublejump}, and this is outside the extended strip for~$s$.
\bigskip

Now that we have established the first four invariants in all cases, we prove Property~\ref{jump-position}, assuming Property~\ref{fill-strip} has been true so far.

Let $t=qb+r$, with $0\le r<b$. Then $qb=\ell$ is the (horizontal) length of the vector 
in $V_t$ that forms the segment of slope~$s$ on $P_t$. We have defined it as
the width of the strip.

To illustrate Property~\ref{fill-strip},
Figure~\ref{fig:one-slope-ordered} shows the possible cases
how the segment of slope $s$ can lie on the grid line, immediately after a jump occurs,
according to this property.
On every grid line, the grid points form an arithmetic progression with
(horizontal)
increment $b$.
The different grid lines are distinguished by the offset $\Delta$ of the leftmost grid point from the left edge of the extended strip.
There are $b$ possibilities, $0\le\Delta<b$.


For the sake of the following analysis, we have sorted the lines by $\Delta$ in
Figure~\ref{fig:one-slope-ordered}. (This is not the order in which they occur from bottom to top.
The true order in this example is $\Delta=0,2,4,1,3,0,\ldots$,
see Figure~\ref{fig:one-slope}.)

For simplicity, we focus on the case when $t$ is even [and put the odd case into brackets].

Let us start with the case $\Delta=0$ (the topmost line in
Figure~\ref{fig:one-slope-ordered}).
In a strip of width $t$, we can fit $q$ segments of length $b$, with $q+1$ points, leaving a remainder of length~$r$.
The extended strip has width $t+\ell$ [$t-1+\ell$].
Since the extra length $\ell$ is filled precisely by $q$ segments,
we can fit $q$ additional segments of length $b$, for a total of $2q+1$ points.
[For odd~$t$, the last claim holds only when $r>0$.]

Since the number of points is odd, the last peeled segment on this line before the jump is a singleton, 
after $q$ steps and at distance $qb=\ell$ from the left boundary $\bar L_s$ of the extended strip, or
distance $\ell-\lfloor t/2\rfloor$ from~$L_s$. 

We can increase $\Delta$ up to $r$ [$r-1$] without changing the situation:

\begin{itemize}
    \item For $\Delta=0,1,\ldots,r$ [$\Delta=0,1,\ldots,r-1$], the number $k$ of points is odd, and for the last point that is peeled, the distance from~$L_s$ is in the range
\begin{displaymath}
    \ell-t/2,\ldots,\ell-t/2+r \quad [
    \ell-\lfloor t/2\rfloor,\ldots,\ell-\lfloor t/2\rfloor+r-1
    ].
\end{displaymath}
Since $\ell+r=t$, this range simplifies to $\ell-\lfloor t/2\rfloor,\ldots,\lfloor t/2\rfloor$.
\end{itemize}

Starting from $\Delta=r+1$ [$\Delta=r$], the situation changes.
We have now an even number $2q$ of points, and
the last peeled segment is a proper segment with a \emph{pair} of points.
The left peeled point is at distance $\Delta+(q-1)b=\Delta+\ell-b$ from
$\bar L_s$, 
or at an offset $\Delta+\ell-b-\lfloor t/2\rfloor$ from~$L_s$ 
(This offset may be negative, in which case it denotes an offset to the left.)
\begin{itemize}
    \item For $\Delta=r+1,\ldots,b-1$ [$\Delta=r,\ldots,b-1$], the number $k$ of points
    is even, and for left point of the last peeled pair,
    the distance from~$L_s$ 
    is in the range
\begin{displaymath}
    r+1+\ell-b-t/2,\ldots,b-1+\ell-b-t/2 \quad 
    [r+\ell-b-\lfloor t/2\rfloor,\ldots,b-1+\ell-b-\lfloor t/2\rfloor].
\end{displaymath}
Since $\ell+r=t$, this range simplifies to $\lfloor t/2\rfloor-b-1,\ldots,\ell-\lfloor t/2\rfloor-1$.
\end{itemize}

Combining the ranges for two cases establishes Property~\ref{jump-position}, and this concludes the proof of Lemma~\ref{predict}.
\end{proof}

\begin{proposition} \label{left-endpoints}
        \label{left-positions}
        The left endpoint of slope $s$ is at distance at most $\lfloor t/2\rfloor$ from $L_s$ (on the left or on the right),
        see Figure~\ref{fig:one-slope-ordered}.
        Every position in this range occurs.
\end{proposition}
\begin{proof}

Property~\ref{jump-position} describes $b$ possibilities before a jump, one value for each of the $b$ residue classes modulo~$b$, and Property~\ref{fill-strip} suggests $b$ possibilities after a jump, namely $\Delta=0,1,\ldots,b-1$. Since for every~$\Delta$, the jump must occur at \emph{some} point, the range \eqref{range-jump} uniquely characterizes this point.
 
By Property~\ref{jump-position}, the left endpoint of slope $s$ can never deviate more than $\lfloor t/2\rfloor$ from $L_s$ to the right, and by Property~\ref{fill-strip}, 
it 
cannot  deviate more than $\lfloor t/2\rfloor$ from $L_s$ to the left. Thus, the left endpoint of slope $s$ is at distance at most $\lfloor t/2\rfloor$ from $L_s$. In fact, since there is a grid line of slope $s$ passing through every such point, and no grid line is skipped (Property~\ref{next-line}), every point in this range will be peeled 
once as the left endpoint of slope $s$.
\end{proof}

There are $2\lfloor t/2\rfloor +1$ different offsets at distance at most $\lfloor t/2\rfloor$ from $L_s$, and exactly one of them is always peeled. Therefore, after $2\lfloor t/2\rfloor +1$ steps the same segment of slope $s$ repeats, one unit higher. This is true for any slope, so after $2\lfloor t/2\rfloor +1$ steps the same chain repeats one unit higher. It means the peeling process is periodic with period $2\lfloor t/2\rfloor +1$, and this concludes the proof of Theorem~\ref{th:reproduce}. \qed

\section{Future research} 

The obvious open problem is to 
prove the relation between grid peeling and
the ACSF for arbitrary convex curves.
As a first challenge, one might try the case of a circle.
The natural approach is to leverage
Theorem~\ref{main-parabola} by locally approximating
the curve by parabolas.


Some of the phenomena that
were revealed by the experiments described in Section~\ref{experiments}
are still awaiting an explanation.


 \bibliographystyle{plainurl}
\bibliography{grid-peeling}

\appendix

\section{Alternative expressions for the horizontal period \texorpdfstring{$H_t$}{H\_t}}
\label{Ht-alternative}

The following long chain of equations and estimates, which we will
discuss step by step,
includes several different expressions for the quantity $H_t$.
We denote by $\mathbb P=\{\,(i,j)\in \mathbb Z\times \mathbb Z \mid
\gcd(i,j)=1\,\}$ the set of \emph{primitive vectors}.
%
  \begin{align}
    H_t 
    &
  := \label{def-Ht}
    \sum_{\substack{  0<y\le x\le t\\
    (x,y)\in \mathbb P
    }}
  \left\lfloor \frac
    tx \right\rfloor x
    \\&
 = \sum_{1\le x\le t} 
    \sum_{\substack{  0<y\le x\\
    (x,y)\in \mathbb P
    }}
  \left\lfloor \frac
    tx \right\rfloor x %
      = \sum_{1\le x\le t} \phi(x)\left\lfloor \frac
    tx \right\rfloor x
    \label{expr-phi}
    \\  &
          =
    \sum_{1\le j\le i\le t} \frac i{\gcd(i,j)}
    \label{expr-gcd}
    \\  &
          =
    \sum_{1\le i\le t}
          \sum_{1\le j\le i} \frac i{\gcd(i,j)}
          \label{expr-gcd2}
    \\&
    =
   \sum_{1\le i\le t} \sum_{d|i} d \phi(d)
    \label{expr-divides}
    \\&
    =
    \frac {2\zeta(3)}{\pi^2}
    t^3 + O(t^2\log t) \label{expr-asymp-precise}
    \\&
    \sim 
 \frac {2\zeta(3)}{\pi^2}
    t^3
    \label{expr-asymp}
  \end{align}
%
  After partitioning the double sum over $x$ and $y$ into two nested sums,
  the last expression in~\eqref{expr-phi} expresses the inner sum, 
  in which the summation variable $y$ does not
  appear in the summand, in terms
  Euler's totient 
  function $\phi(x)$, 
  the number of residue
  classes $y$ modulo $x$ that are relatively prime to~$x$.

The asymptotic expression \eqref{expr-asymp}
is due to
Sándor and Kramer \cite{sandor99}. They define the function that they
investigate by the expression~\eqref{expr-gcd2}, using the notation
$\psi_1(i)$ for the inner sum in \eqref{expr-gcd2},
and they establish
equality between \eqref{expr-gcd2} 
and the last expression in~\eqref{expr-phi} 
\cite[\S6, formula~\thetag{13}
with $\alpha=1$. Formula~\thetag{13$'${$'$}$'$}, which should apply here, has an obvious typo]{sandor99}.
The expressions  \eqref{expr-gcd} and \eqref{expr-gcd2}
are obviously equivalent.
We indicate why 
\eqref{expr-gcd}
is equal to~\eqref{def-Ht}, thus
establishing equality of all expressions \thetag{\ref{def-Ht}--\ref{expr-gcd2}}:

In the sum \eqref{def-Ht},
only primitive vectors appear, but
  every primitive vector $(x,y)\in \mathbb{P}$ is
taken with multiplicity $\lfloor t/x\rfloor$.
If we look at the multiples $(i,j)=
(kx,ky)$ of each primitive vector $(x,y)$ 
for $k=1,2,\ldots,\lfloor t/x\rfloor$, we get \emph{all} integer vectors
in the triangle $0<j\le i\le t$.
In the sum~\eqref{expr-gcd},
 each vector $(i,j)$ contributes
the $x$-coordinate of
the primitive vector $(x,y)=(i/k,j/k)$, since $k=\gcd(x,y)$. 
Thus, in total, the vector $(x,y)$ is taken
$\lfloor t/x\rfloor$ times in
\eqref{expr-gcd}, and we
get the same sum.

Formula~\eqref{expr-divides} follows
from~\eqref{expr-gcd2}
by splitting the integers $j=1,\ldots i$
according to the value of $i/\gcd(i,j)=d$, cf.~\cite[\S2, formula~\thetag{1}]{sandor99}.

The asymptotic expression \eqref{expr-asymp}
is given in \cite[equation~\thetag{23}, p.~61]{sandor99}. 
A different proof, using the Wiener-Ikehara
Theorem, was sketched by Marko Riedel on Mathematics Stack
Exchange~\cite{riedel14}.
The slightly stronger statement~\eqref{expr-asymp-precise}
with the explicit 
error term
follows from Lemma~\ref{vectors-in-triangle}.

\section{Proof of Lemma~\ref{vectors-in-triangle} about the points on the grid parabola}
\label{proof-alpha}
\input{asymptotic-triangle}

\section{Minimum-area convex lattice polygons and grid parabolas}
\label{min-area}
\begin{figure}
    \centering
    \includegraphics[scale=0.9]{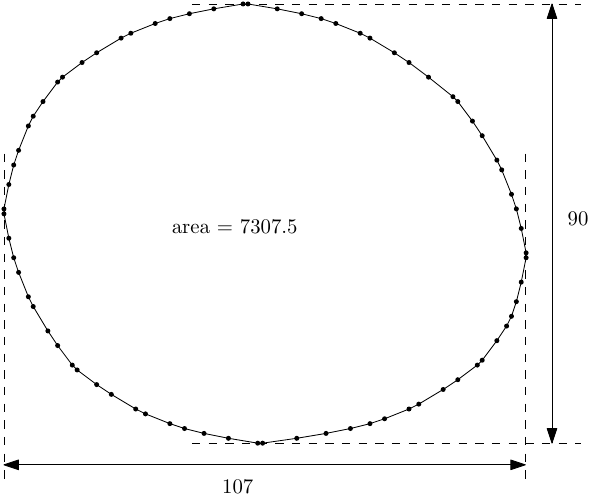}
    \caption{The lattice 75-gon with smallest area}
    \label{fig:75-gon}
\end{figure}

\begin{figure}
    \centering
    \includegraphics{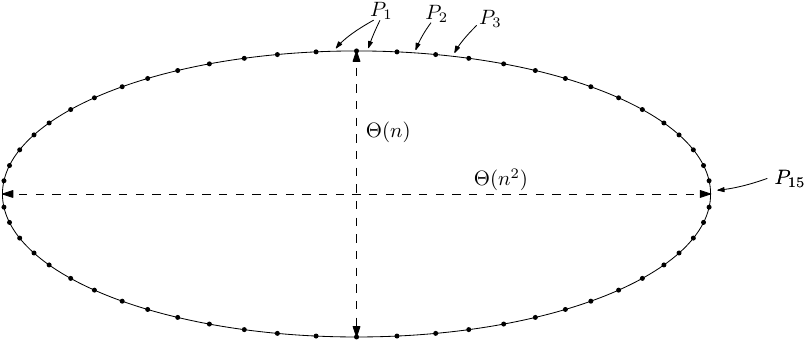}
    \caption{As $n$ increases, the minimum-area lattice $n$-gon becomes more and more oblong.}
    \label{fig:min-area-ellipse}
\end{figure}

Our grid parabolas $P_t$ make their appearance in a different context:
the convex lattice $n$-gons of minimum area for a given number $n$ of vertices,
see Figure~\ref{fig:75-gon} for an example.
Bárány and 
Tokushige~\cite{bar-tokushige-20}
showed that, as $n$ increases, the smallest $n$-gon resembles a more and more oblong ellipse-like shape,
whose ``axes'' grow like $n^2$ and $
n$, respectively,
see the schematic drawing in Figure~\ref{fig:min-area-ellipse}.
It follows from their analysis (although this is not explicitly stated) that, after
a unimodular transformation,
the optimal $n$-gons are composed of pieces of the grid parabolas $P_1,P_2,\ldots,P_m$ with horizontal axes.
There is a global finite bound $m$ on the number pieces.
There is strong numerical evidence that $m=15$, like shown in the picture, and thus, every such polygon consist of (at most)
58 pieces, taken from the  grid parabolas $P_1,P_2,\ldots,P_{15}$.

\section{Experiments with grid peeling for parabolas}
 \label{experiments}

Our work was initiated by experimentally exploring the grid peeling
process for parabolas $\Pi\colon y=ax^2+bx$, i.e.,
parabolas with a vertical axis.
\ifanonymous\else
These experiments were carried out
in the
M.\,Ed.~thesis of Moritz R\"uber~\cite{rueber-thesis}, and we
report some of the findings from this thesis.
\fi
We tried different rational values of~$a$, sometimes in combination with
various values of $b$, and started the grid peeling process.
We let the process run until it reached a curve that was
a translate of a previous curve, and the process became periodic.
Then we could estimate the average vertical speed as
well as various other quantities.

In order to carry out these experiments, we needed to restrict
the computations to a finite range, 
as mentioned in the introduction,
and therefore we
had to compute the horizontal period~$H$: The smallest horizontal translation that, in combination with a shearing operation, maps both the parabola $\Pi$ and the integer grid to itself.
%
The following lemma,
which is analogous
to Lemma~\ref{lem:period-Pt}
for the grid parabola,
gives a formula for $H$:

\begin{figure}
    \centering
    \includegraphics{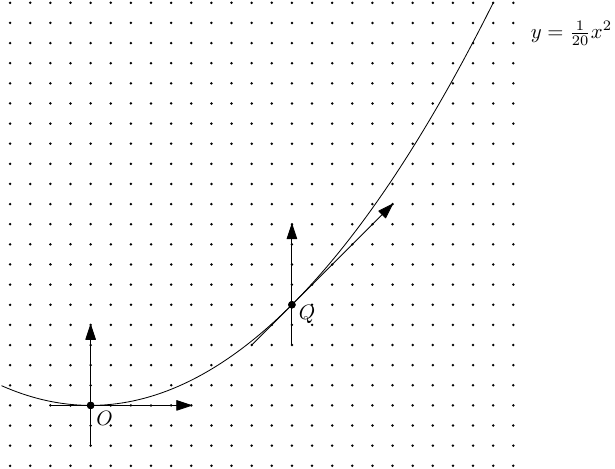}
    \caption{An affine grid-preserving transformations maps the origin $O$ to the point $Q$ and it maps the parabola to itself.
    The horizontal period $H$ is 10 in this example.}
    \label{fig:one-parabola}
\end{figure}

\begin{proposition} \label{prop:horizontal-period-parabola}
  Consider the parabola
\begin{equation} \label{parabola}
    y=\frac{a_N}{a_D}x^2 +
  \frac{b_N}{b_D}x + c     
 \end{equation}
with reduced fractions $\frac{a_N}{a_D}$ and $\frac{b_N}{b_D}$,
and let $\bar H := \lcm(a_D,b_D)$.
 
  The horizontal period $H$ is either equal to
  $\bar H$ or to
  $\bar H/2$. In particular
  \begin{enumerate}
  \item
[(i)]
    If $a_D\equiv 0 \pmod 4$, then $H = \lcm(a_D/2,b_D)$.
  \item [(ii)]
 If $a_D\equiv 2 \pmod 4$
and $b_D\equiv 2 \pmod 4$,
    then $H = \lcm(a_D/2,b_D/2)$.
  \item [(iii)]
In all other cases,  $H = \lcm(a_D,b_D)$.
  \end{enumerate}
  
\end{proposition}

  \begin{proof}
  See Figure~\ref{fig:one-parabola} for an example.
  Since vertical lines should remain vertical
  and vertical distances should remain unchanged,
  the affine transformation that we are looking for has
  the form
  \begin{equation}\label{transformation} 
      \begin{pmatrix}x\\y
      \end{pmatrix}
      \mapsto
      \begin{pmatrix}\hfill x&&{\!\!\!\!\!}+H\hfill\\
      zx&{\!\!\!\!\!}+y &{\!\!\!\!\!}+w
      \end{pmatrix}
  \end{equation}
  for suitable parameters $H,z,w$.
    Substituting the right-hand sides of \eqref{transformation} for $x$ and $y$
    should leave the parabola equation \eqref{parabola} unchanged.
 This leads to the following equations for $z$ and~$w$:
    \begin{align}
      \label{eq:z}
      z &= 2\tfrac{a_N}{a_D}H
\\
      \label{eq:w}
      w &= \tfrac{a_N}{a_D}H^2 + \tfrac{b_N}{b_D}H 
    \end{align} 
The additional requirement is that
\eqref{transformation} should map integer grid points to
integer grid points. This boils down to requiring that
 $H,z$, and $w$ are integral.

Thus, the horizontal period that we are looking for
is the smallest positive integer $H$ for which the two quantities $z$
    and $w$ defined above
    are integers.

Obviously, the choice $H = \bar H := \lcm(a_D,b_D)$ makes both
 \eqref{eq:z} and
\eqref{eq:w} integral.
The remainder of the proof, which is elementary but goes into
case distinctions,
only amounts to checking whether some
smaller value of $H$ 
     also does the job.
     
We start with an easy observation:
    
 Observation A:
    If $H$ and $z$ are integers and $z$ is even, then
    $w$ is an integer
 iff  $H$ is a multiple of $b_D$.

This can be seen 
after
rewriting the first term of \eqref{eq:w},
using \eqref{eq:z} to express one factor $H$ in terms of $z$,
    \begin{equation}
      \label{eq:w2}
      w = \tfrac{zH}{2} + \tfrac{b_N}{b_D}H 
    \end{equation}
 If $zH$ is even, then
  the first term $\tfrac{zH}{2}$ in  \eqref{eq:w2} is an integer.
Therefore, $w$ is an integer iff the second term
is also an integer, which is the case iff $H$ is a multiple of $b_D$.

We make a case distinction based on the parity of $a_D$:
    
Case 1: $a_D$ is even: Then,
from \eqref{eq:z},
$z=\frac{a_N}{a_D/2}H$ is an integer iff $H$ is a multiple of
$a_D/2$.

Case 1.1: $a_D$ is a multiple of 4: Then $H$ is even, and
Observation A implies that
$w$ is an integer iff $H$ is a multiple of $b_D$.
Part~(i) follows.

Case 1.2:
$a_D\equiv 2 \pmod 4$.
Then $a_N$ must be odd,
because $\tfrac{a_N}{a_D}$ is reduced.

We distinguish two subcases, depending on the (unknown) value $H$.

Case 1.2.1: $H$ is odd. In this case,
$z=\frac{a_NH}{a_D/2}$ is also odd, and
the first term $\tfrac{zH}{2}$ in  \eqref{eq:w2} is a half-integer
(a non-integer multiple of $1/2$).
For $w$ to be an integer, the second term
$\tfrac{b_NH}{b_D}$
must also be a half-integer. This is only possible if the factor 2 is
contained in the denominator $b_D$. Then $b_N$ must be odd,
because $\tfrac{b_N}{b_D}$ is reduced. If $b_D$ were divisible by 4,
then, since the numerator $b_NH$ is odd, 
$\tfrac{b_NH}{b_D}$ could not be a half-integer.
Hence the only remaining possibility is 
$b_D\equiv 2 \pmod 4$.
Then
$\tfrac{b_NH}{b_D}$ is a half-integer iff
$\tfrac{b_NH}{b_D/2}$ is an integer, which holds iff
$H$ is a multiple of
$b_D/2$.
The value $\lcm(a_D/2,b_D/2)=\bar H/2
$ is odd,
and therefore it
is indeed the smallest viable value of $H$ in this case.

Case 1.2.2: $H$ is even.
Then $H$
must be a multiple of $\lcm(2,a_D/2)= a_D$.
Observation~A implies that
$w$ is an integer iff $H$ is also a multiple of $b_D$.
Thus $H$ must be a multiple of $\bar H=\lcm(a_D,b_D)$.

Summing up the two subcases, we see that
the value $\bar H/2$ for Case 1.2.1 is indeed the smallest
value of $H$ when this case is possible, i.e., when
$b_D\equiv 2 \pmod 4$. This
establishes Part~(ii) of the Proposition.
When
Case 1.2.1 does not apply, the value $H=\bar H$
is in accordance with Part~(iii).

Case 2: $a_D$ is odd: Then $z$ is an integer iff $H$ is a multiple of
$a_D$, and $z$ will be even.
As above, Observation A implies that
$w$ is an integer iff $H$ is also a multiple of $b_D$.
The smallest possible value $H$ is
$\bar H=\lcm(a_D,b_D)$, and
this is in accordance with Part~(iii).    
  \end{proof}



\begin{figure}
    \centering
    \includegraphics[scale=0.9]{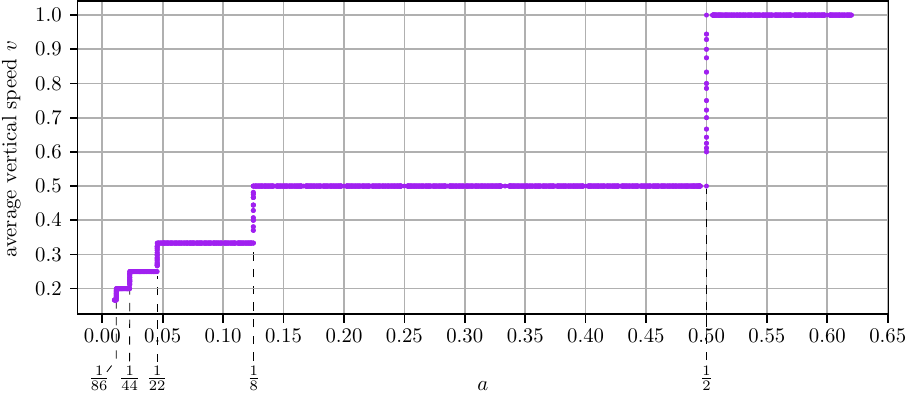}
    \caption{
The average vertical speed $v$ of the
parabola $y=ax^2+bx$
versus the coefficient $a$ (for various values of~$b$).}
    \label{fig:average-vertical-speed}
\end{figure}

\subsection{Results of the experiments}

Parabola $y=ax^2+bx$ for various rational values $a$ and $b$.
We computed the convex hull of the grid points above the parabola
and started to peel.
After some preperiod, the peeling process will settle into a cyclic behavior:
After a certain number $\Delta m$ of steps, the same curve reappears, translated upward by $\Delta y$. We call
$\Delta m$ the \emph{time period} and
$\Delta y$ the \emph{vertical period}. The \emph{average (vertical) speed}
$v$ is then 
$\Delta y/\Delta m$. 
Figure~\ref{fig:average-vertical-speed}
shows the
average vertical speed depending on $a$ (for various values of $b$).

These experiments show a very clear picture.
For almost all
~$a$, 
the average speed takes one of 
the values $1,\frac12,\frac13,\ldots$, and it does not depend on $b$.
The sharp transitions between the values occur at 
\emph{critical values} of $a$, which we can recognize, with
hindsight, as $1/(2H_t)$. In fact, this is how the phenomenon described in 
Section~\ref{sec:grid-parabola}
was discovered: The sequence of denominators of
the critical values $\frac 12, \frac{1}{8}, \frac{1}{22},$ etc.,
after clearing the common factor 2, appears in the
O.E.I.S.~
\cite[\href{https://oeis.org/A174405}{A174405}]{OEIS},
where 
various formulas are given, including
\eqref{expr-gcd2}
and~\eqref{expr-divides} in Section~\ref{Ht-alternative}.
By trying to find an 
interpretation of these formulas that would make sense
in the context of a convex curve, we were 
led to the discovery of the grid parabolas~$P_t$.
Experiments soon revealed their remarkable behavior under grid
peeling, as described in Theorem~\ref{th:reproduce}.

\subsection{The average speed at the critical values of \texorpdfstring{$a$}{a}}
\label{b-at-critical}

At the critical values $a=\frac1{2H_t}$,
the  average vertical speed varies between
the consecutive fractions
$\frac 1{t+1}$ and
$\frac 1{t}$.
\begin{figure}
    \centering
    \includegraphics[scale=0.8]
    {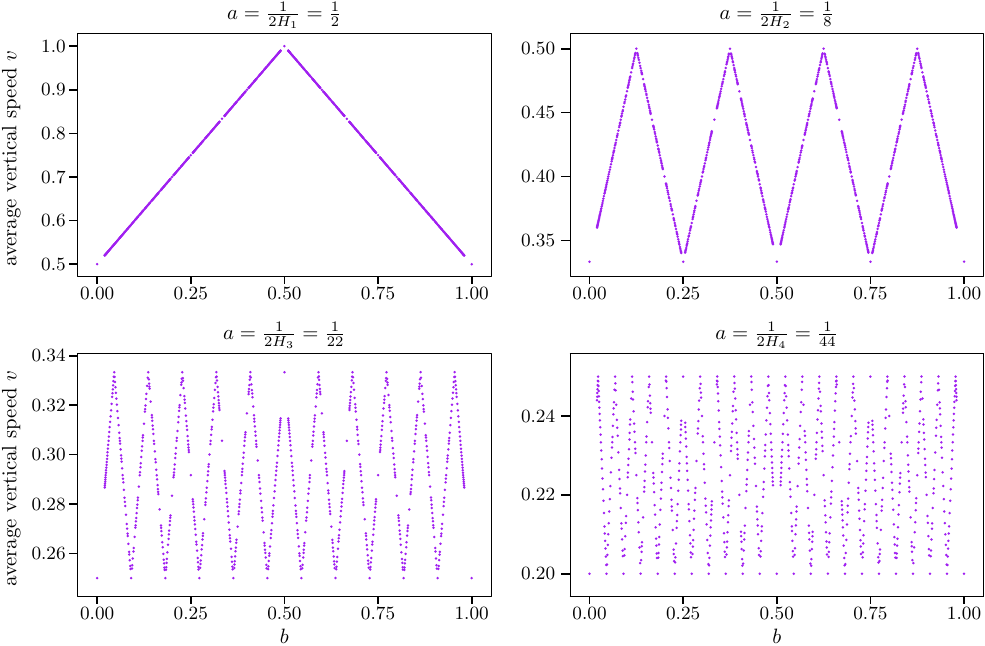}
    \caption{Average vertical speed for critical values $a=\frac1{2H_t}$
    for $t=1,2,3,4$,
    depending on $b$}
    \label{fig:critical}
\end{figure}
Figure~\ref{fig:critical} shows the average vertical speed for the
first four critical values~$a$, depending on~$b$.
We see a very regular, piecewise linear dependence on $b$,
filling the range between
$\frac 1{t+1}$ and
$\frac 1t$.

The horizontal periodicity of these graphs can be explained easily:
\begin{proposition}\label{period-b}
    The parabolas $y=ax^2+bx$
and
    $y=ax^2+b'x$ with $b'=b+
    2a$ 
    have the same average speed.
\end{proposition}

\begin{figure}
    \centering
    \includegraphics{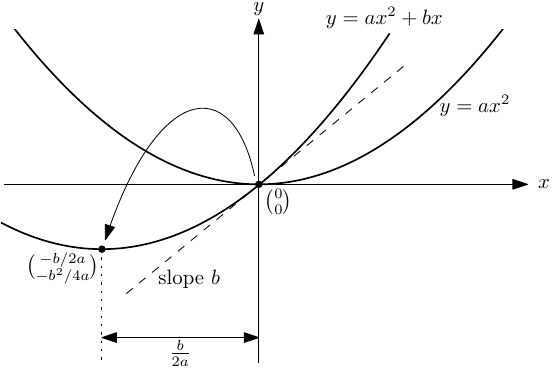}
    \caption{
    Changing the parameter $b$ of the parabola $y=ax^2 + bx$ amounts to a translation.    }
    \label{fig:vertical-vs-horizontal-shift}
\end{figure}
\begin{proof}
    See Figure~\ref{fig:vertical-vs-horizontal-shift}.
The ``shape'' of the parabola depends only on $a$.
Variation of $b$ can be interpreted as a translation,
placing the vertex of the parabola to a different location.
More precisely, the vertex of $y=ax^2+bx$ lies at the minimum
of the function, $x=-\frac{b}{2a}$.
Thus, increasing $b$ by 
$2a$ incurs an
amount of horizontal translation  that
is integral.
By Proposition~\ref{invariant-translations},
the resulting parabola has then the same average speed.
\end{proof}

We see that the graph 
consists of $H_t$ repetitions
of a sawtooth.
Proposition~\ref{period-b} explains the periodicity with period $1/H_t$.
It is also clear 
exchanging $b$ with $-b$ performs only a reflection at
the $y$-axis and thus
has no effect on the average vertical speed. This mirror symmetry, together with the periodicity with period $1/H_t$ implies
that the whole graph is determined by
the part in the interval  $0\le b\le \frac1{2H_t}$.

What remains is the remarkable fact
that the
average speed grows linearly in this interval.
We have no explanation for this phenomenon.

The acute observer may 
notice irregular gaps in
the dotted lines of Figure~\ref{fig:critical}. These gaps are,
however, only an artifact of the way
how the values $b$ were chosen: We took all reduced fractions $b=\frac{b_N}{b_D}$ with $0\le b_N\le b_D\le 50$. Such a set
leaves 
gaps around values with small denominator
like $1, \frac12, \frac13, \frac23$, etc.,
while it fills other intervals more densely.
(In Figure~\ref{fig:time-period} below,
this choice of parameters is also the reason why
the lower parts of the figure do not
exhibit the periodicity of Proposition~\ref{average-speed},
but rather show the data points arranged along various curves.)

\subsection{The time period at the critical values of \texorpdfstring{$a$}{a}}
\begin{figure}
    \centering
    \iflong
    \includegraphics[scale=1.05]{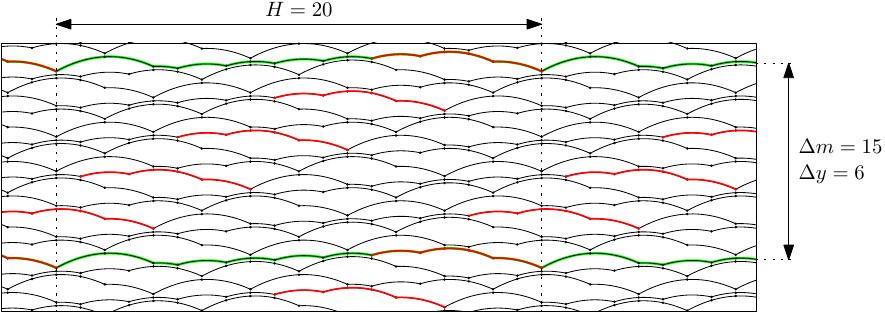}
    \else
    \hskip-3mm
    \includegraphics[scale=0.95]{period-with-shift.pdf}
    \fi
    \caption{Grid peeling of the parabola
    $\Pi\colon y = \frac18x^2 + \frac 15x
    = \frac1{2H_t}x^2 + \frac 15x
    $ for $t=2$, after 
    the periodic behavior has started.
    Each curve $C$ is regarded as a function $C(x)$ of $x$,
    and we draw the function $C(x)-(\frac18x^2 + \frac 15x)$ instead
    of $C(x)$.
    Straight segments have turned into 
    downward-curving parabolic arcs.    
    The original parabola $\Pi$ would appear as
    a horizontal line.
    (This is the same drawing convention that was used
    for the grid parabolas in
    Figure~\ref{fig:diff-reference}.)
    }
    \label{fig:period-with-shift}
\end{figure}

Note that an average speed like $0.4=2/5$ (as for the parabola $y = \frac18x^2 + \frac 15x$, for example) 
means
that we can no longer have a vertical period 
of 1: To get
the fraction $2/5$,
the vertical period $\Delta y$ must be a multiple of 2, and
the time period $\Delta m$ must be a multiple of 5.
In this example, the true periods are 
$\Delta y=6$ and $\Delta m=15$, see Figure~\ref{fig:period-with-shift}.
We can see that the curve reappears already after $3$ iterations,
combined with a horizontal shift by $4$ units.
Some characteristic pieces of the curves are marked in red in
 Figure~\ref{fig:period-with-shift} to highlight
 this repetition.
Note that the appearance as a \emph{translation}
is due
to the special drawing style of Figure~\ref{fig:period-with-shift},
which makes also the horizontal period $H$
appear as a 
translational symmetry.
What appears as a translation is
actually 
an affine transformation.

 The horizontal period
 is $H=20$, and thus, after $20/4=5$ shifts (and
  $5\times 3
  =15$  
 iterations), the
 shifted curve coincides with the original curve, completing
 a cycle.

\begin{figure}[htb]
    \centering
    \iflong
    \includegraphics[scale=0.97]{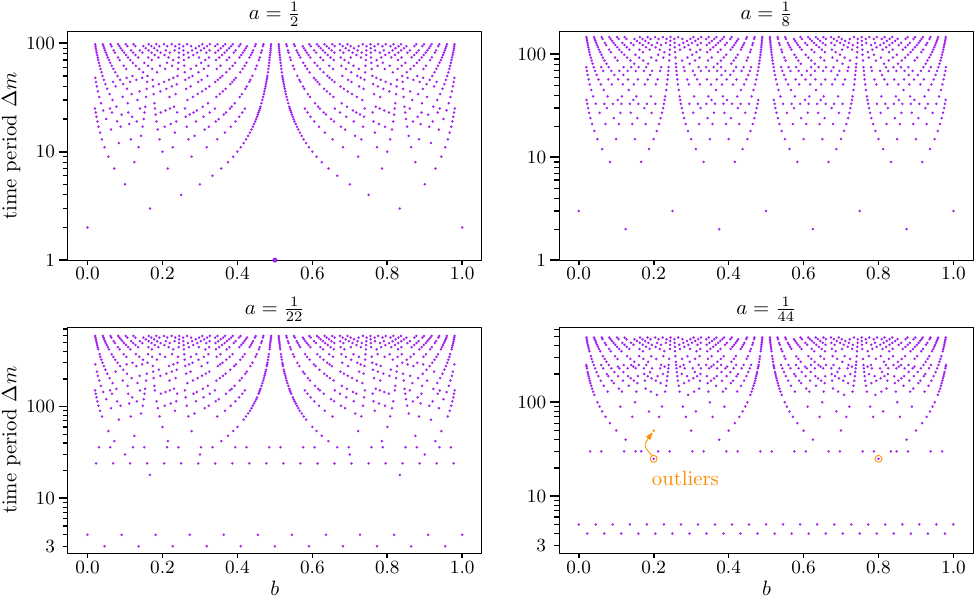}
    \else
    \noindent\hskip-6mm
    \includegraphics[scale=0.87]{time-period2.pdf}
    \fi
    \caption{Time period at the critical values $a$, as a function of $b$.
    The orange point in the graph for $a=\frac{1}{44}$ shows where the
    value should be if formula~\eqref{predict-Delta-m} were true.}
    \label{fig:time-period}
\end{figure}

 This pattern seems to hold for most examples:
 The curve returns after a ``subperiod'' of $\binom {t+1}2$ steps, shifted by
 a multiple of $H_t$, and all
 multiples of $H_t$ modulo $H$ appear as shifts.
 Thus, the time period $\Delta m$, under these assumptions,
 is equal to
 \begin{equation}\label{predict-Delta-m}
 \Delta m = \binom {t+1}2 \frac{H}{H_t} =
 \frac{t(t+1)H}{2H_t},
 \end{equation}
where the horizontal period $H$ is given by
Proposition~\ref{prop:horizontal-period-parabola}.
Figure~\ref{fig:time-period} shows the time period
$\Delta m$ for the same set of parabolas
as Figure~\ref{fig:critical}, on a logarithmic
scale.

For the cases when $b$ is a multiple of $a$,
where the vertical speed takes the ``extreme'' values
$\frac 1{t+1}$ or
$\frac 1t$, the time period is $t+1$ or $t$ respectively,
with $\Delta y=1$.
For all remaining cases with one exception,
the estimate \eqref{predict-Delta-m} predicts the correct
value. The exception is the parabola
$y=\frac{1}{44}x^2 + \frac{1}{5} x$
and its symmetric twin
$y=\frac{1}{44}x^2 + \frac45 x$
with
$\Delta m=25$,
whereas
 \eqref{predict-Delta-m} would give $\Delta m=50$.
The reason behind this exceptional behavior is that
the curve reappears already after $5$ iterations (with
a horizontal shift of $44$) instead of $\binom{t+1}{2}=\binom{5}{2}=10$
iterations.
We suspect that such exceptions will turn up more frequently
when more values of $b$ and higher values of~$t$ are tested.

In any case, the experiments revealed 
interesting 
patterns in the grid peeling process for
parabolas
(extending the results stated for grid parabolas in Theorem~\ref{th:reproduce} and Lemma~\ref{predict}),
whose precise structure remains to be discovered and analyzed. 

We tried to see whether the grid parabola $P_t$ appears when
peeling is started with an appropriate parabola.
To obtain the time period predicted by 
Theorem~\ref{th:reproduce},
as suggested by the data of Figure~\ref{fig:time-period},
we tried the parabolas
$y=\frac{1}{2H_t}x^2 + \frac{1}{2H_t}x+\gamma$ for odd $t$
and
$y=\frac{1}{2H_t}x^2 +\gamma$ for even $t$,
introducing some vertical shift~$\gamma$.
These parabolas have also the correct horizontal period $H=H_t$
by Proposition~\ref{prop:horizontal-period-parabola},
using the fact that $H_t$ is even iff $t$ is even%
\iflong
\ 
(Proposition~\ref{prop:parity}.\ref{parity}).
\else
.
(The last statement can be proved along the lines of
Proposition~\ref{prop:parity}:
the vectors that are matched in a pair sum to a vector with an even $x$-coordinate. The unmatched vectors in 
 $V_t^{(0)}$ are the vector of slope $s=\frac12$, which has an even $x$-coordinate,
 and the vector $\binom tt$, whose parity therefore decides the
 parity of~$H_t$.)
\fi

We found that, although the average speed does not change with $\gamma$,
the periodic cycle which the process enters can change.
For $t\le 5$, we always found $P_t$ among one of the periodic cycles,
with an appropriate choice of $\gamma$.

If the piecewise linear dependence of the average speed
that is shown by
the experiments holds for all values of~$b$, this means that there must
be parabolas with an irrational average speed, namely when $b$ is irrational.
This would imply that the peeling process is necessarily aperiodic
(a fact that can of course not be confirmed experimentally).

\subsection{Deviation from the parabolic shape
}

For a curve $C$ that arises in the peeling process of $\Pi$ we can compute
the deviation from the shape of the starting parabola $\Pi$ as follows:
We find the highest translate $\Pi+\gamma$ that lies below~$C$
and the lowest translate $\Pi+\gamma'$ that lies above $C$, such
that $C$ is contained in a parabolic tube of (vertical)
thickness 
$\gamma'-\gamma$.

\begin{figure}[htb]
    \centering
    \includegraphics[scale=0.8]{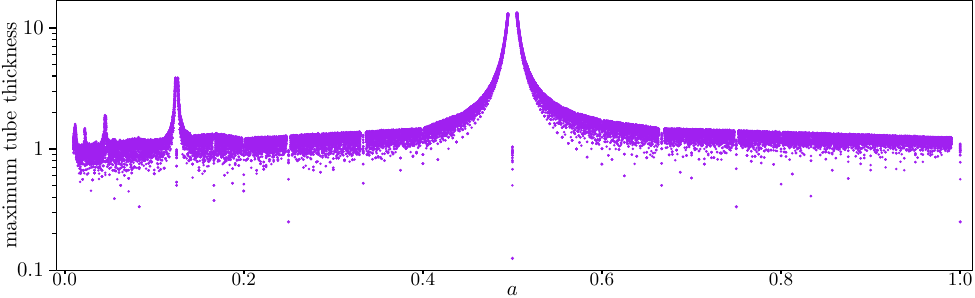}
    \caption{The tube thickness versus the coefficient $a$,
    for various values of~$b$.}
    \label{fig:maxtube}
\end{figure}

We looked at all parabolas $y=ax^2+bx$ with
rational coefficients $a$ and $b$ in the interval
$0\le a,b <1$, where the denominator of $a$ ranges between 1 and 100, and the denominator of $b$ between 1 and 10.
For each parabola,
we observed 
 the maximum tube thickness of all curves arising
  during the periodic part of the peeling process.
Figure~\ref{fig:maxtube} plots the maximum tube thickness
on a logarithmic scale and
the parameter $a$. The dependence on $b$ is not shown.

Most of the time, the tube thickness is near 1.
Interestingly, it rises to high values
when $a$ is close to one of the critical values $\frac{1}{2H_t}$,
seemingly approaching a vertical asymptote.
At the critical values themselves, the tube thickness is, however, lower than usual. 


\section{
Vertical difference between the grid parabola and the reference parabola}
\label{appendix-difference}

\begin{figure}
    \centering
    \iflong
    \includegraphics{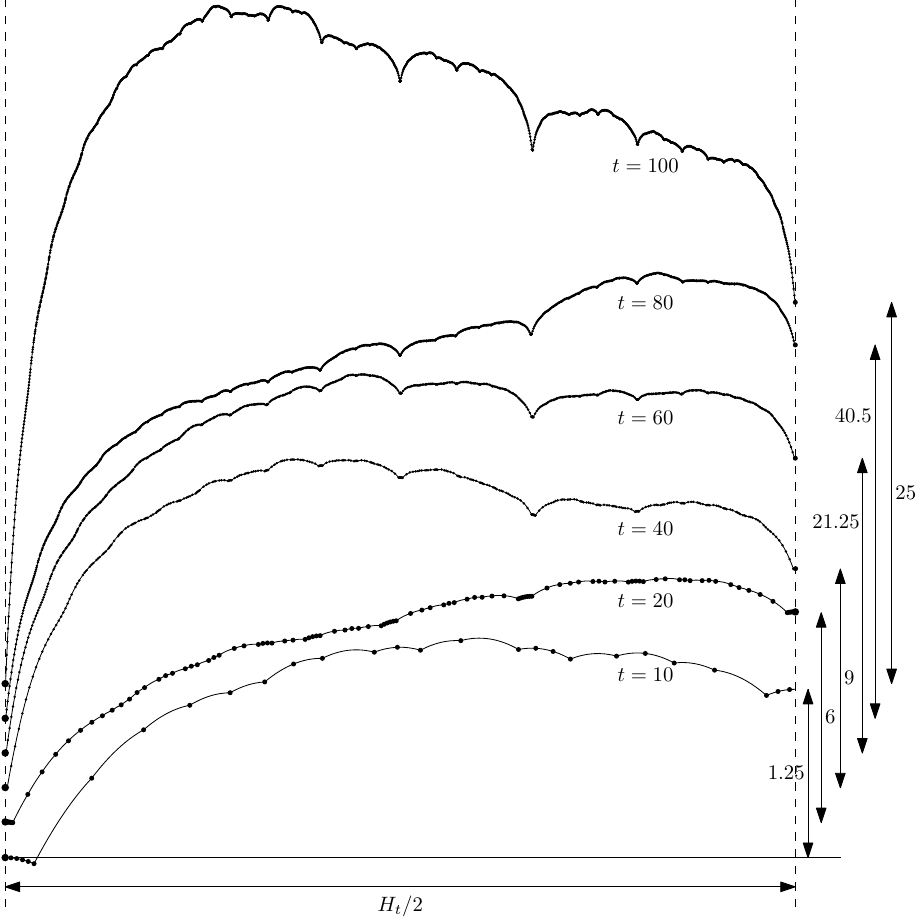}
    \else
    \includegraphics[scale=0.895]{diff-reference2.pdf}
    \fi
    \caption{Vertical difference $P_t-\Pi_t$ for $t=10,20,40,60,80,100$. 
    The horizontal axis ranges from the origin to the point $x=H_t/2$, where 
    both $P_t$ and $\Pi_t$ have slope~$\frac12$.
    The boundaries of this interval, marked by dashed lines,
    act as vertical mirrors for
    the function $P_t-\Pi_t$.
    The curves are vertically shifted so that they can be nested
    under each other,
    and the vertical scale is chosen independently for each curve: 
    The vertical double-headed arrows show the range between 
    0 and the difference at  $x=H_t/2$. 
    (The outlier value $40.5$ for $t=80$ is no mistake. The curve for $t=80$ rises indeed to a higher maximum than for $t=100$.)
    }
    \label{fig:diff-reference}
\end{figure}

For some selected values of $t$,
we have computed the vertical difference $P_t-\Pi_t$  as a function of~$x$,
and we show the result in
Figure~\ref{fig:diff-reference}.

We observe that $P_t$ is always above $\Pi_t$, with the exception of the horizontal segment of length $t$ around
the origin.
We can clearly discern
local minima near $H_t/2$, $H_t/3$ and $H_t/4$, and less marked minima near
other rational multiples of $H_t$.

\end{document}

%% file: asymptotic-triangle.tex

The proof follows standard ideas; in particular, it
uses an easy adaptation of the arguments that were used
for the special case $\alpha=1$ \cite{sandor99},
as stated in \iflong\else the second equation of \fi Lemma~\ref{lem-Ht-asymptotic}. 
%
Following \cite[p.~61]{sandor99},
we reduce the computation of $U_t^\alpha$ to
the estimation of the vector~$R^\alpha(u)$, which is defined as follows
for any real parameter $u$: 
\begin{align}
 R^\alpha(u) &:=
        \label{def-R-alpha}
    \sum_{\substack{ 1\le x\le u
    \\0<y\le \alpha x\\
    (x,y)\in \mathbb P}}
  \binom xy
\end{align}
We start from the definition of $U_t^\alpha$
and regroup and reformulate sums:
\begin{align}
  U_t^\alpha 
         \nonumber
  &=
    \sum_{\substack{  1\le x\le t \\ 0<y\le \alpha x \\(x,y)\in \mathbb P}}
  \left\lfloor \frac
  tx \right\rfloor \binom xy \\
  \nonumber
  &=
    \sum_{\substack{1\le x\le t \\ 0<y\le \alpha x \\(x,y)\in \mathbb P}} \,
  \sum_{\substack{k\ge 1\\ xk\le t}} \binom xy
                                             =
    \sum_{{k\ge 1}} 
 \,   \sum_{\substack{ 1\le x\le t \\xk\le t\\ 0<y\le \alpha x \\(x,y)\in \mathbb P}}
  \binom xy
          =
    \sum_{\substack{k\ge 1\\\ }}\,
    \sum_{\substack{ 1\le x\le t/k\\ 0<y\le \alpha x\\(x,y)\in \mathbb P}}
    x\\
  &=
    \sum_{k\ge 1} R^\alpha(\tfrac tk) \label{X-to-R-alpha}
\end{align}
For $R^\alpha(u)$, we will use the following asymptotic estimate,
whose proof is given below:
\begin{lemma}\label{lemma-R-alpha}
For $0\le\alpha\le1$,
  \begin{equation}\label{R-asymp-alpha}
    R^\alpha(u) =
     \frac {2}{\pi^2}
     \binom{u^3\alpha+O(u^2\log u)}{u^3\alpha^2/2+O(u^2\log u)}
\end{equation}
as $u\to\infty$.
\end{lemma}

Substitution of
\eqref{R-asymp-alpha}
in \eqref{X-to-R-alpha} gives our claimed asymptotic formula for $U_t^\alpha$:
\begin{align*}
  U^\alpha_t = \sum_{k\ge 1} R^\alpha\bigl(\tfrac tk\bigr) 
        &= \sum_{k\ge 1} \frac 2{\pi^2 k^3}
        \binom{t^3\alpha}{t^3\alpha^2/2}
+ \sum_{k\ge 1} \binom
{O\bigl(\tfrac{t^2}{k^2}\log \tfrac tk \bigr)}
{O\bigl(\tfrac{t^2}{k^2}\log \tfrac tk \bigr)}
  \\&
  = \frac {2\zeta(3)}{\pi^2} 
  \binom{t^3\alpha}{t^3\alpha^2/2}
  + \binom
  {O(t^2\log t)}
  {O(t^2\log t)}
\end{align*} 


 \begin{proof}[Proof of Lemma~\ref{lemma-R-alpha}]

We mention that the special case $\alpha=1$ and the $x$-coordinate of
$R^1(u)$
is given  (without the error bound) in
Sándor and Kramer \cite[p.~61, \thetag{20}]{sandor99}:
  \begin{equation}\nonumber 
    \sum_{\substack{ 1\le y\le x\le u\\
    (x,y)\in \mathbb P}}
  x
  =
     \sum_{1\le x \le u} 
    \sum_{\substack{ 1\le y\le x\\
    (x,y)\in \mathbb P}}
  x
  =
     \sum_{1\le x \le u} x\phi(x)
     \sim
     \frac {2u^3}{\pi^2}
\end{equation}
There, this asymptotic expression is
derived as an easy consequence
of a general statement of Radoux~\cite{radoux77} about the sum
$\sum_{x=1}^u f(\tfrac xu)\phi(x)$ being asymptotically equal to
${6u^2}\!/{\pi^2}\!\int_{z=0}^1zf(z)\,dz$ for an arbitrary function $f$,
provided that $zf(z)$ is continuous.  (The deeper reason behind this
statement is that the fraction of primitive vectors among the integer
vectors in \emph{any} sufficiently large ``well-behaved'' region is
$1/{\zeta(2)}$.
%
Examples of well-behaved regions for which this statement holds
are convex regions with a limit on the ratio between incircle and
circumcircle, or dilates of a fixed convex region.)

The following 
proof
of~Lemma~\ref{lemma-R-alpha}
adapts the textbook derivation of the similar formula
\begin{equation}
  \nonumber
    \sum_{\substack{ 1\le y\le x\le u\\(x,y)\in \mathbb P}}
  1
=    \sum_{x=1}^{\lfloor u\rfloor}\phi(x)
=\frac {u^2}{2\zeta(2)}+O(u\log u) \sim
\frac {3u^2}{\pi^2}
\end{equation}
in \cite[Theorem~330]{HW}, a result that goes back to Mertens in 1874.
%



   We use the notation $T^\alpha(u)$ for the sum
\eqref{def-R-alpha}
   without the condition that
     $(x,y)$ should be primitive:
\begin{align}
 T^\alpha(u) &:=
        \sum_{\substack{1\le x\le u
        \\1\le y\le \alpha x
        }}  \binom xy
        \label{def-T-alpha}
 \\  
      &=    \sum_{x=1}^{\lfloor u\rfloor}
      \binom{x\lfloor \alpha x\rfloor}
      {\lfloor \alpha x\rfloor(\lfloor \alpha x\rfloor+1)/2}
  \label{with.rounding}
 \\  
      &=    \sum_{x=1}^{\lfloor u\rfloor}\binom
      {\alpha x^2 + O(x)}
      {\alpha^2 x^2/2 + O(x)}
 \label{without.rounding}
      \\&
  =
  \binom{\alpha}{\alpha^2/2}
  \left(   \frac{ \lfloor u\rfloor^3}3 +\frac{\lfloor u\rfloor^2}2 +
  \frac{\lfloor u\rfloor}6 \right)
  + \binom{O(u^2)}{O(u^2)}
  \label{u2}
  \\  &
                           = \binom{\alpha}{\alpha^2/2}
                           \frac {u^3}3 + \binom{O(u^2)}{O(u^2)}
                           \label{T-approx-alpha}
\end{align}
 We claim that the $O(u^2)$ error terms
 in \eqref{T-approx-alpha} can be explicitly bounded by
\begin{equation}
  \label{bound-E-alpha}
  \left\lVert T^\alpha(u)-
  \binom{\alpha}{\alpha^2/2} \frac{u^3}3\right\rVert_\infty \le
  E(u) :=
  \begin{cases}
  \frac{u^3}3  & \text{for $0\le u<1$} \\
    \frac{5u^2}3  & \text{for $u\ge 1$} \\
  \end{cases}
\end{equation}
The bound
for the first case, $u< 1$, is trivial because $T^\alpha(u)=\binom00$ in this case.
For the second case, $u\ge 1$,
we accumulate the error bounds for the successive steps of the above derivation.
The $O(x)$ error term in the
transition from \eqref{with.rounding} to \eqref{without.rounding}
is bounded (in absolute value) by $x$ in the first 
coordinate and by $\alpha x/2\le x
$ in the
second 
coordinate.
Thus, we can bound each of the $O(u^2)$ terms in \eqref{u2} by
$ u(u+1)/2\le u(2u)/2 = u^2$.

The change in the factor between the term $A(u)
:=\frac{\lfloor u\rfloor^3}3+ \frac{\lfloor u\rfloor^2}2+\frac {\lfloor
      u\rfloor}6
$
 in~\eqref{u2}
 and the factor $u^3/3$ in
 \eqref{T-approx-alpha} is bounded by $2u^2/3$, as
can be checked by an easy computation:
\begin{align*}
A(u)&= \textstyle 
\frac{\lfloor u\rfloor^3}3+ \frac{\lfloor u\rfloor^2}2+\frac {\lfloor
      u\rfloor}6
      \le
      \frac{u^3}3+ \frac{u^2}2+\frac u6
      \le
      \frac{u^3}3+ \frac{u^2}2+\frac {u^2}6
  \\
A(u)&= \textstyle 
\frac{\lfloor u\rfloor^3}3+ \frac{\lfloor u\rfloor^2}2+\frac {\lfloor u\rfloor}6 \ge
\frac{(u-1)^3}3+ \frac{(u-1)^2}2+\frac {u-1}6
=\frac{u^3}3- \frac{u^2}2+\frac {u}6
      \ge
      \frac{u^3}3- \frac{u^2}2-\frac {u^2}6.
\end{align*}
and therefore,
$|A(u)-\frac{u^3}3|\le 
    \frac{2u^2}3$.
Adding the two contributions together gives the claimed bound of 
$u^2+2u^2/3 = 5u^2/3$.

Comparing
\eqref{def-R-alpha}
with the sum~\eqref{def-T-alpha}, for which we have an explicit formula, we
want to exclude
the vectors $(x,y)$ that are not primitive, i.e., where both $x$ and $y$ are
 multiples of one of the primes $2,3,5,7,11,\ldots$.
By the inclusion-exclusion formula,
we have to
subtract the contribution of those vectors that are divisible by
 $2$, by 3, by 5, etc., add the vectors that are divisible by \emph{two}
 primes, subtract the vectors that are divisible by three primes, etc.
The contribution of the vectors that are divisible by $n$ is $nT^\alpha(\frac
un)$, as these vectors are all integer vectors from the smaller region
$0< x\le u/n, \, 0\le y\le\alpha x$, scaled by~$n$.
We obtain:
 \begin{align*}
   R^\alpha(u) &= T^\alpha(u) - \bigl(2T^\alpha(\tfrac u2) + 3T^\alpha(\tfrac u3) + 5T^\alpha(\tfrac u5) + \cdots
          \bigr)\\
   &\qquad
   + 
     \bigl(2\cdot 3\cdot  T^\alpha(\tfrac u{2\cdot 3}) +
     2\cdot 5\cdot  T^\alpha(\tfrac u{2\cdot 5}) +
     \cdots \bigr)
\\   &\qquad
   -
   \bigl(2\cdot 3 \cdot 5\cdot  T^\alpha(\tfrac u{2\cdot 3\cdot 5}) + \cdots \bigr) + \cdots
   \\
   &= \sum_{n=1}^\infty \mu(n) n T^\alpha(\tfrac un),
 \end{align*}
 In the last line, we have expressed the alternating sum
 in terms of the Möbius function
 \begin{displaymath}
   \mu(n) =
   \begin{cases}
     +1, & \text{if $n$ is the product of an even number of distinct primes,}\\
     -1, & \text{if $n$ is the product of an odd number of distinct
       primes,}\\
     0, & \text{otherwise, i.e., if $n$ is not square-free.}
   \end{cases}
 \end{displaymath}
 We now use the approximation
 \eqref{T-approx-alpha} for $T^\alpha(\frac un)$
\begin{equation}
  \label{H-mu-alpha}
  R^\alpha(u)   = \sum_{n=1}^\infty \mu(n) n T^\alpha(\tfrac un)=
            \sum_{n=1}^\infty \mu(n) \frac {u^3}{3n^2}
            \binom{\alpha}{\alpha^2/2}+ E_0
\end{equation}
 and bound the error
 $E_0$ by
\eqref{bound-E-alpha}: 
 \begin{align*}
  \lVert E_0\rVert_\infty
  &\le
        \sum_{n=1}^\infty \left|\mu(n) n E(\tfrac un)\right|
  \le
        \sum_{n=1}^\infty  n E(\tfrac un)
   \\
 &\le
        \sum_{n=1}^{\lfloor u\rfloor} \frac {5u^2}{3n}
 +
   \sum_{n=\lfloor u\rfloor+1}^\infty  \frac {u^3}{3n^2}
   =\frac{5u^2}3 \cdot O(\log u) + \frac{u^3}3\cdot O(
   \tfrac1u)
   =O(u^2\log u) 
 \end{align*}
The first term in \eqref{H-mu-alpha} can be expressed in terms of the
zeta-function
using the well-known relation \cite[Theorem~287]{HW}
(which also derives from the inclusion-exclusion formula)
\begin{align*}
  \sum_{n=1}^\infty \frac {\mu(n)}{n^2}
  &
    =
    \biggl(1-\frac1{2^2}\biggr)
    \left(1-\frac1{3^2}\right)
    \left(1-\frac1{5^2}\right) 
    \left(1-\frac1{7^2}\right) \cdots
  \\
  &=
    \frac1{1\!+\!\tfrac1{2^2}\!+\!\tfrac1{4^2}\!+\!\tfrac1{8^2}\!+\!\cdots}
   \cdot \frac1{1\!+\!\tfrac1{3^2}\!+\!\tfrac1{9^2}\!+\!\tfrac1{27^2}\!+\!\cdots}
   \cdot \frac1{1\!+\!\tfrac1{5^2}\!+\!\tfrac1{25^2}\!+\!\tfrac1{125^2}\!+\!\cdots}
    \cdots\\
  &=
    \frac1{\sum_{n=1}^\infty\frac1{n^2}} = \frac1{\zeta(2)} = \frac 6{\pi^2}
.
\end{align*}
We obtain
\begin{displaymath}
    R^\alpha(u)    = \sum_{n=1}^\infty \mu(n) 
    \frac {u^3}{3n^2}
    \binom{\alpha}{\alpha^2/2} + E_0
 = \frac {6u^3}{3\pi^2}\binom{\alpha}{\alpha^2/2}  + E_0
 = \frac {2}{\pi^2}\binom
 {u^3\alpha+O(u^2\log u)}{u^3\alpha^2/2+O(u^2\log u)} 
 .
 \qedhere
\end{displaymath}  
 \end{proof}